%% file: paper.tex
\documentclass[conference]{IEEEtran}

\usepackage[font=footnotesize,caption=false]{subfig}
\PassOptionsToPackage{hyphens}{url}
\usepackage{xcolor}
\usepackage{hyperref}
\hypersetup{
  colorlinks,
  linkcolor={green!65!black},
  citecolor={red!70!black},
  urlcolor={blue!70!black}
}

\usepackage{graphicx}
\graphicspath{ {figs/} }

\usepackage{amsmath}
\usepackage{amsfonts}
\usepackage{amssymb}
\usepackage{mathtools}

\let\labelindent\relax
\usepackage[binary-units=true]{siunitx}
\DeclareSIUnit\packet{p}
\DeclareSIUnit\query{q}
\newcommand{\percent}[1]{\SI[round-precision=1,round-mode=places]{#1}{\percent}}

\usepackage{csquotes}
\usepackage[
    capitalise,
    nameinlink,
]{cleveref}

\usepackage[inline, shortlabels]{enumitem}
\usepackage{changepage}
\newenvironment{indended}
    {\vspace{.5\baselineskip}\begin{adjustwidth}{1cm}{}}
    {\end{adjustwidth}\vspace{.5\baselineskip}}

\newcommand{\CNAME}{} 
\def\CNAME/{\texttt{CNAME}}
\newcommand{\DNSKEY}{} 
\def\DNSKEY/{\texttt{DNSKEY}}
\newcommand{\NS}{} 
\def\NS/{\texttt{NS}}
\newcommand{\A}{} 
\def\A/{\texttt{A}}
\newcommand{\SRV}{} 
\def\SRV/{\texttt{SRV}}

\newcommand{\ip}[1]{%
    {\fontfamily{lmtt}\selectfont%
        #1%
    }%
}
\newcommand{\rfc}[1]{%
RFC\,#1~\cite{rfc#1}%
}

\usepackage{acronym}
\makeatletter
\newcommand*{\org@overidelabel}{}
\let\org@overridelabel\@verridelabel
\@ifpackagelater{acronym}{2015/03/21}{
    \renewcommand*{\@verridelabel}[1]{%
        \@bsphack
        \protected@write\@auxout{}{\string\AC@undonewlabel{#1@cref}}%
        \label{#1}%
        \org@overridelabel{#1}%
        \@esphack
    }%
}{
    \renewcommand*{\@verridelabel}[1]{%
        \@bsphack
        \protected@write\@auxout{}{\string\undonewlabel{#1@cref}}%
        \label{#1}%
        \org@overridelabel{#1}%
        \@esphack
    }%
}

\newcommand{\acposs}[1]{%
    \expandafter\ifx\csname AC@#1\endcsname\AC@used
    \acs{#1}'s%
    \else
    \aclu{#1}'s (\acs{#1})%
    \fi
}
\makeatother

\newacro{authns}[AuthNS]{Authoritative Name Server}
\newacro{cdn}[CDN]{Content Delivery Network}
\newacro{doh}[DoH]{DNS over HTTPS}
\newacro{dot}[DoT]{DNS over TLS}
\newacro{fpr}[FPR]{False Positive Rate}
\newacro{isp}[ISP]{Internet Service Provider}
\newacro{knn}[$k$-NN]{$k$-Nearest Neighbors}
\newacro{pir}[PIR]{private information retrieval}
\newacro{sni}[SNI]{Server Name Indication}
\newacro{tpr}[TPR]{True Positive Rate}
\newacro{wf}[WF]{Website Fingerprinting}
\usepackage{etoolbox}

\begin{document}

\date{}

\title{Padding Ain't Enough: Assessing the\\Privacy Guarantees of Encrypted DNS}

\author{\IEEEauthorblockN{Jonas Bushart}
\IEEEauthorblockA{CISPA Helmholtz Center for Information Security\\
jonas.bushart@cispa.saarland}
\and
\IEEEauthorblockN{Christian Rossow}
\IEEEauthorblockA{CISPA Helmholtz Center for Information Security\\
rossow@cispa.saarland}}

\maketitle

\pagestyle{plain}

\input{abstract}

\input{introduction}
\input{background}
\input{threatmodel}
\input{methodology}
\input{evaluation}
\input{countermeasures}
\input{discussion}
\input{relatedwork}
\input{conclusion}

\IEEEtriggeratref{54}

\clearpage
\bibliographystyle{IEEEtranS}
\bibliography{references,ietf-rfcs,urls}

\clearpage
\input{appendix}

\end{document}

%% file: abstract.tex
\begin{abstract}
\ac{dot} and \ac{doh} encrypt DNS to guard user privacy by hiding DNS resolutions from passive adversaries.
Yet, past attacks have shown that encrypted DNS is still sensitive to traffic analysis.
As a consequence, RFC~8467 proposes to pad messages prior to encryption, which heavily reduces the characteristics of encrypted traffic.
In this paper, we show that padding alone is insufficient to counter DNS traffic analysis.
We propose a novel traffic analysis method that combines size and timing information to infer the websites a user visits purely based on encrypted \emph{and padded} DNS traces.
To this end, we model \emph{DNS sequences} that capture the complexity of websites that usually trigger dozens of DNS resolutions instead of just a single DNS transaction.
A closed world evaluation based on the Alexa top-10k websites reveals that attackers can deanonymize at least half of the test traces in \percent{80.2} of all websites, and even correctly label all traces for \percent{32.0} of the websites.
Our findings undermine the privacy goals of state-of-the-art message padding strategies in \acs{dot}/\acs{doh}.
We conclude by showing that successful mitigations to such attacks have to remove the entropy of inter-arrival timings between query responses.
\end{abstract}

%% file: introduction.tex
\section{Introduction}
\label{sec:introduction}

The Domain Name System (DNS) is the inherent starting point of nearly all Internet traffic, including sensitive communication.
However, despite its vast popularity, the DNS standards explicitly do not address well-known privacy concerns.
As a consequence, in traditional DNS, passive adversaries can easily sniff on name resolutions, and for example, trivially identify the Web services that users aim to visit.
Eavesdroppers can use this information for advertisement targeting, or to create nearly-perfect profiles of the browsing/usage behavior of their clients.
While we seem to have accepted this lack of DNS privacy for years, recently, the IETF standardized two new protocols that aim to protect user privacy in DNS.
Both proposals, \acf{dot}~\cite{rfc7858} and \acf{doh}~\cite{rfc8484}, wrap the DNS communication between DNS client and resolver in an encrypted stream (either TLS or HTTPS) to hide the client's name resolution contents from passive adversaries.

These privacy-preserving protocols have seen rapid and wide adoption.
Many prominent open DNS resolvers started to support \ac{dot} or \ac{doh}, such as Cloudflare, Quad9, or Google Public DNS.
Likewise, several popular DNS resolver implementations (e.g., Bind, PowerDNS, Unbound) offer \ac{dot} and/or \ac{doh} functionality~\cite{dns-privacy-implementation-status}.
Similarly, these new standards have been adopted on the client side.
Most prominently, Google deployed \ac{dot} for all Android~9 devices~\cite{android-dot}, and Mozilla started to support \ac{doh} in Firefox~62~\cite{firefox-doh}.
Furthermore, vendors have announced support in more popular clients for the near future, such as in Chrome~\cite{chrome-doh-cl}.
These developments towards encrypted DNS will help to preserve privacy (and integrity) of DNS communication, and thus have the potential to close a long-lived, staggering privacy gap.

Yet, similar to other privacy-preserving communication systems that leverage encryption (e.g., HTTPS~\cite{Schuster2017,Wright2009} or Tor~\cite{Panchenko2016,Sirinam2018,Wang2014}), \ac{dot} and \ac{doh} are susceptible to traffic analysis.
Gillmor's empirical measurements~\cite{ndss-padding} show that passive adversaries can leverage the mere size of a single encrypted DNS transaction to narrow down the queried domain.
Such size-based traffic analysis attacks were expected in the according standards, and it is thus not surprising that the IETF seeked for countermeasures.
Most prominently, \rfc{8467} follows Gillmor's suggestions to pad DNS queries and responses to multiples of 128~bytes and 468~bytes, respectively.
It is widely assumed that this padding strategy is a reasonable trade-off between privacy and communication overhead~\cite{ndss-padding}.
In fact, popular \ac{dot}/\ac{doh} implementations (e.g., BIND, Knot) and open resolvers (e.g., Cloudflare) heavily rely on these current best practices, although still in experimental state.

In this paper, we study the privacy guarantees of this widely-deployed \ac{dot}/\ac{doh} padding strategy.
We assume a passive adversary (e.g., \acs{isp})---notably not the DNS resolver itself---who aims to deanonymize the DNS resolutions of a Web client.
We follow the observation that modern Web services are ubiquitously intertwined.
When visiting a website, clients leave DNS traces that go beyond individual DNS transactions.
While the suggested padding strategy destroys the length entropy of individual messages, we assess to what extent DNS resolution \emph{sequences} (e.g., caused by third-party content embedded into Web services) allow adversaries to reveal the Web target.
Such sequences not only reflect characteristic series of DNS message sizes, but also expose timing information between DNS transactions within the same sequence.

Based on this information, we propose a method to measure the similarity between two DNS message sequences.
We then leverage a \acf{knn} classifier to search for the most similar DNS transactions sequences in a previously-trained model.
We leverage this classifier to assess if attackers can reveal the website a used visited purely by inspecting the DNS message sequence.
Our closed world evaluation based on the Alexa top-10k websites reveals that attackers can deanonymize five out of ten test traces in \percent{80.2} of all websites, and correctly label all traces for \percent{32.0} of the websites.
We measure the false classification rate in an open world scenario and show that attackers can limit the false classification rate to \percent{10} while retaining true positive rates of \percent{40}.
Even classifying DNS sequences that exclude assumed-to-be-cached DNS records can preserve the classifier's accuracy if such partially cached sequences are included in the training dataset.
Overall, our results show that adversaries can reveal the content of encrypted \emph{and padded} DNS traffic of Web clients, which so far was believed to be impossible.

Our findings undermine the privacy goals of state-of-the-art message padding strategies in \ac{dot}/\ac{doh}.
This is highly critical, especially when seen in light of recent orthogonal initiatives to anonymize Web services, ranging from \emph{HTTPS Everywhere} to encrypting domains in \acs{sni}~\cite{ietf-tls-esni-02}).
Admittedly, DNS is not the only---and with \ac{dot}/\ac{doh} certainly not the easiest---angle to reveal the communication target.
In fact, IP addresses of endpoints, traffic analysis of HTTPS traffic~\cite{Schuster2017,Wright2009}, and even plaintext domain information in TLS's \acs{sni} are (currently) easier targets to reveal the Web services a client visits.
However, \ac{dot}/\ac{doh} were designed with clear privacy guarantees in mind, and it is crucial that we have a thorough understanding whether these promises can be kept up.

Seeing this loss in privacy, we thus discuss countermeasures against sequence-based DNS traffic analysis.
We show that even an optimal (i.e., overly aggressive) padding strategy does \emph{not} provide sufficient privacy guarantees.
Instead, we find that countermeasures also have to reduce the entropy of inter-arrival timings between query responses.
We thus propose to extend padded encrypted DNS with countermeasures that mitigate timing side channels.
In particular, we show that both constant-rate sending and Adaptive Padding~\cite{Shmatikov2006} preserve DNS privacy.
Our prototypical implementation demonstrates their effect as a trade-off for bandwidth and timing overheads.

Summarizing, we provide the following contributions:
\begin{enumerate}
    \item We illustrate a traffic analysis attack that leverages DNS transaction \emph{sequences} to reveal the website a client visits, as opposed to existing deanonymization attempts that consider only individual transactions and can easily be mitigated by message padding.

    \item We provide an extensive analysis of the privacy guarantees offered by DNS message padding (\rfc{8467}) against our attack in a Web browsing context.
    We demonstrate severe privacy losses even against passive adversaries that sniff on encrypted and padded DNS traffic only.

    \item We evaluate alternative padding strategies and constant-rate communication systems against our proposed attack.
    We reveal that even perfect padding cannot mitigate traffic analysis, and show that any promising countermeasure will have to lower the entropy of timing information.
\end{enumerate}

%% file: background.tex
\section{Background}
\label{sec:background}

DNS is one of the most fundamental protocols on the Internet.
As its primary use case, DNS allows online services to expose domain names that can easily be remembered by clients and which, upon request, will be translated into IP addresses.
Upon visiting a website, users query DNS resolvers, typically even for multiple domain names to load third-party Web content.
The DNS resolver either has the answer cached, or performs an iterative lookup by relaying the query to \acp{authns}, which can then answer the DNS lookup.
The response then flows back from the \ac{authns} to the DNS resolver, subsequently, to the user/client.

In this work, we focus on name resolutions, DNS's predominant use case.
In general, DNS also forms the basis for several other key ingredients of the Internet.
There are many email anti-spam methods, such as DKIM~\cite{rfc6376}, SPF~\cite{rfc7208}, DNS-based blacklists~\cite{dnsbl}, TLS certificate pinning (DANE)~\cite{rfc7671,rfc6698}, and even identity provider protocols (ID4me)~\cite{id4me}.

\subsection{Security Assessment of DNS}
\label{ssec:secofdns}

For years, DNS deployments did not foresee any means to protect the integrity and confidentiality of DNS messages.
The lack of integrity has allowed for manipulation attempts, such as malicious captive portals or redirecting non-existing domains to malicious websites.
Worse, without integrity protection, an attacker could alter the DNS resolution for critical websites, such as banks, and redirect visitors to launch phishing attempts.
By now, the increasing deployment of DNSSEC~\cite{dnssec-validation-world,dnssec-deployment} allows an \ac{authns} to cryptographically sign DNS responses to prevent that attackers can spoof or manipulate responses before they reach the DNS resolver.

While DNSSEC ensures message integrity, it explicitly does \emph{not} provide confidentiality guarantees.
As a consequence, in plain DNS (even with DNSSEC), both queries and responses leak the requested domain names.
This has severe privacy implications, such as leaking the browsing history of users to any party who can observe the DNS communication, such as \acp{isp}, attackers sniffing on unencrypted Wi-Fi traffic, or other on-path attackers.

\subsection{Encrypted DNS}
\label{ssec:active-proposals}

Users are understandably worried about privacy in DNS-based name resolutions.
One particular critical path is between the client and the resolver, where several entities such as the \ac{isp} might sniff on the communication.
To close this gap, in the last years, various proposals evolved to encrypt the DNS communication between clients and recursive resolvers~\cite{dnscrypt,rfc8484,huitema-quic-dnsoquic-05,rfc7858,rfc8094}.
The common element of all proposals is to encrypt and authenticate the DNS traffic to guarantee both, client-to-resolver integrity \emph{and} confidentiality.
Among these approaches there are two predominantly deployed proposals, which mostly differ in their transport protocol:

\textbf{\acf{dot}} is specified in \rfc{7858} from March 2016.
    It takes the existing DNS over TCP protocol and wraps it into a TLS layer.
    DNS over TCP is specified as part of the original DNS RFCs, \rfc{1035}, and is supported by all major DNS resolver implementations.
    Supporting \ac{dot}, besides adding a TLS stack, requires only minor changes to DNS implementations.
    The specification also allows for an opportunistic encryption mode, in which the identity of the DNS server cannot be verified, but encryption is still used, thus providing protection against passive network attackers.
    \ac{dot} is well-suited for a system-wide DNS service such as the stub resolvers of operating systems.
    For example, Android~9 introduces a system-wide \ac{dot} resolver~\cite{android-dot}.

\textbf{\acf{doh}} is a more recent proposal from October 2018 and is specified in \rfc{8484}.
    It contains two different approaches, based on the HTTP GET and POST methods.
    The GET method uses a URI with a parameter in the form of \texttt{\nolinkurl{https://dnsserver.example/?dns=<DNS_in_base64>}} where \enquote{\texttt{<DNS\_in\_base64>}} represents a base64 encoding of a DNS message in DNS wire format.
    In the case of HTTP POST, the DNS request is part of the HTTP message body in DNS wire format.
    Regardless of the request method, the DNS reply is sent in the HTTP message body in wire format.
    As of time of writing, this is the only standardized response format, but \rfc{8484} allows for future extensions.
    Given its HTTP focus, interest in \ac{doh} stems largely from browser vendors.
    As such, both Firefox~62+~\cite{firefox-doh} and Chromium-based browsers already support \ac{doh}~\cite{bromite,chrome-doh-cl}.

\subsection{Threat Model of \acs{dot}/\acs{doh}}
\label{ssec:proposal-scope}

Encrypted DNS aims to guarantee confidentiality under the assumption that the resolver is trusted.
Both \ac{dot} and \ac{doh} secure the communication between a DNS client and the DNS resolver only.
This threat model aims to protect against on-path attackers such as \acp{isp}, Wi-Fi access points, or Internet Exchange Points.
Trusting the DNS resolver is essential.
As the DNS resolver still sees all requests, it can trivially break confidentiality and integrity.
The user thus has to carefully choose a resolvers with a privacy policy they like and trust.
While it does not protect privacy, to guarantee integrity, clients can perform DNSSEC validation on top of encrypted DNS.

Note that all upstream communication of the resolver (towards \acp{authns}) is not covered by \ac{dot}/\ac{doh} and remains plain DNS.
The \ac{isp} of the DNS resolver and any upstream on-path attacker (between resolver and \acp{authns}) can still see the iterative lookups.
Yet this privacy impact is less severe, as resolvers mix the queries of many clients giving some weak notion of group anonymity, which makes it difficult to map clients to queries.
In addition, the resolvers cache responses of all their clients globally, and thus not every client query will be visible to the \ac{isp}.
In this paper, we will follow this threat model and assume a passive attacker that can observe the communication between a DNS client and its resolver.

%% file: threatmodel.tex
\section{Problem Statement and Threat Model}
\label{sec:threat-model}

Encrypting DNS communication, as in \acs{dot} and \acs{doh}, strongly increases user privacy.
Yet: \emph{Which privacy guarantees do users of these new standards obtain with regard to potential side channels?}
In this work, we investigate how much an attacker can learn by side-channel analysis on encrypted and padded DNS traffic.
Following the threat model of encrypted DNS schemes, our honest-but-curious attacker passively sniffs (TLS/HTTPS-encapsulated) DNS traffic between the client and its resolver to infer which websites the user visits.
Such an attacker might be an \ac{isp}, a VPN hoster, or a Wi-Fi operator trying to monetize user profiles.
Our attacker does not delay, alter, or inject data into the user's traffic.
However, they may perform stateful analysis.
Finally, we assume that the attacker can initiate their own network connections from the same network-topological location as the attack victim---a trivially satisfied assumption in the Wi-Fi and \ac{isp} setting.
This final assumption enables the attacker to create a training dataset that accurately represents the client's perspective.

Our concrete attack scenario is that we investigate if encrypted and padded DNS helps to hide which website a user visits.
We assume a passive adversary that can investigate encrypted and padded DNS traffic between the client and its resolver only.
Following the threat model of \ac{dot}/\ac{doh} (\cref{ssec:proposal-scope}), we explicitly define trivial attacks such as sniffing on the resolver's upstream communication as out of scope.
Furthermore, we assume that the user picks a trusted DNS resolver that does not breach user privacy.
We choose Web browsing as the one use case of our privacy analysis, as it is the most common use case for casual computer usage.

%% file: methodology.tex
\section{Methodology}
\label{sec:methodology}

We now present how traffic analysis can infer the browsing target purely based on encrypted DNS traffic.
We start by discussing if traffic analysis can still infer message contents despite the fact that the encrypted DNS messages are padded.
Then, we present \emph{DNS sequences}, a new data structure that derives features from encrypted DNS traffic.
Finally, we propose a \ac{knn} classifier that uses a custom distance metric that reveals which website visit has caused a given DNS sequence.

\subsection{Traffic Analysis and DNS Padding}
\label{ssec:traffic-analysis-and-dns-padding}
Encrypted DNS does not leave many characteristics that can be leveraged to infer the communication content.
In fact, we can observe features in three dimensions, namely
\begin{enumerate*}[(i)]
    \item counts, such as packets,
    \item sizes, such as overall transmitted bytes, and
    \item time.
\end{enumerate*}
Naturally, combinations of these characteristics are possible, e.g., bursts, which measure the packets per time interval.
These three dimensions---despite encryption---may provide valuable hints about the communication content.
For example, the strings \enquote{en.wikipedia.org} and \enquote{bit.ly} are easily discernible by their length.
To increase privacy, one can thus reduce the aforementioned dimensions.
Most importantly, \rfc{7830} provides a mechanism to pad the size of DNS messages to a fixed length, by adding a new extension type to EDNS~\cite{rfc6891}.
\rfc{8467} recommends padding all queries to a multiple of \SI{128}{\byte}, while all responses are padded to a multiple of \SI{468}{\byte}.
This way, padding largely unifies byte counts and reduces one dimension for traffic analysis.
\rfc{8467} follows the work of Gillmor~\cite{ndss-padding}, who empirically studied the effects of various padding strategies based on a sample of recorded DNS traffic.
Based on this, \rfc{8467} suggests a padding strategy that is a trade-off between bandwidth overhead and the risk of privacy violations.
\textbf{Several popular \ac{dot}/\ac{doh} implementations (e.g., BIND, Knot) and public resolvers (e.g., Cloudflare\footnote{\url{https://dnsprivacy.org/jenkins/job/dnsprivacy-monitoring/} (see \enquote{Padding})}) thus already rely on this padding strategy, and we expect that others will follow.}

\subsection{DNS Sequences}
\label{ssec:dns-sequence}
Gillmor chose the padding recommendations after studying the privacy implications for \emph{individual} query-response pairs.
However, as we will show, attackers can leverage a \emph{sequence} of DNS query/response pairs instead, which heavily increases the uniqueness compared to individual DNS transactions.
These sequences of DNS queries occur for different reasons during Web browsing, such as redirects, loading of third-party resources, or resources from subdomains, all of which can only happen after the initial GET request, thus forming a dependency between DNS queries.
As we will show, these sequences characterize a website a user visited quite well.

We thus use a \emph{DNS sequence} data structure as the core representation to perform traffic analysis.
Such a DNS sequence abstractly captures sizes (Msg) and inter-arrival times (Gap) of DNS messages.
Before going into details, we will use an example to illustrate the basics of the underlying idea.
Consider a visit to \texttt{\nolinkurl{wikiquote.org}} which triggers four DNS requests/responses upon visit:
\begin{enumerate*}[(i)]
	\item at the start: \texttt{\nolinkurl{wikiquote.org}}
	\item after \SI{287}{\ms}: \texttt{\nolinkurl{www.wikiquote.org}}
	\item and after \SI{211}{\ms}: both \texttt{\nolinkurl{meta.wikimedia.org}} and \texttt{\nolinkurl{upload.wikimedia.org}} simultaneously.
\end{enumerate*}
The resulting DNS sequence (which ignores requests) looks like this:

\begin{indended}
	Msg(1), Gap(8), Msg(1), Gap(7), Msg(1), Msg(1)
\end{indended}

All responses are \SI{468}{\byte} large, and there are \SI{287}{\milli\second} and \SI{211}{\milli\second} time between them, respectively.
The DNS sequence encodes the four DNS responses (each \SI{468}{\byte} long, i.e., exactly one padding block).
If there is a time gap between two responses, we note this gap and its magnitude.
To be less susceptible to timing variations due to typical jitter, we express the time in log scale (e.g., $\lfloor log_2(287) \rfloor = 8$) instead of accurate numbers.
There is no Gap between the last two Msgs, as they arrive (almost) simultaneously.
That is, we neglect all time gaps with a numerical value $\leq 0$ (i.e., those shorter than \SI{1}{\milli\second}).

We choose to focus on downstream DNS communication (DNS replies) only.
That is, we extract our features from the data stream from the DNS resolver to the client only.
There are several reasons why we chose this design.
First, in a test of popular domains, based on the Alexa domain list~\cite{alexa}, we found that \emph{all} DNS requests are sufficiently small to be padded to the same size of \SI{128}{\byte}.
This is the smallest padding size for a request and it shows that requests provide little entropy.
Second, the timing of requests is also highly correlated with the timing of the responses, as most queries will be cached by the DNS resolver.
Thus, one round-trip time after the query, the response can be seen on the wire.
Furthermore, in the face of reordering, it is impossible to reliably correlate responses to queries in case of multiple (equally-looking) parallel queries.
Overall, our focus on replies simplifies the design, while it retains the traffic characteristics.

In order to represent message sizes and timestamps, we opted for an abstract notion of size and time.
This provides higher flexibility to generalize different implementations and events outside of our control (e.g., network performance, jitter).
Technically, instead of exact response sizes, we are only interested in the padding group the message belongs to, i.e., which multiple of \SI{468}{\byte} was used for padding.
Furthermore, we compute the differences in the message timestamps to create a coarse-grained estimate of how long the gap to the previous DNS response was.
Decreasing the granularity of times increases robustness against changing timings, e.g., due to different bandwidth limits, network conditions, or hardware configurations.
We thus infer the rounded logarithmic time differences between two consecutive DNS replies:
$$
gap_i(t_{i-1}, t_i) = \lfloor log_2(t_i - t_{i-1}) \rfloor.
$$

Knowing the DNS message sizes and their time gaps, both of which can be extracted from encrypted and padded DNS traffic, enables us to create a DNS sequence that characterizes a website.
Formally, a DNS sequence is a list of $n$ elements $(e_1, \dotsc, e_n)$.
An element $e_i$ is part of the set $\{\text{Msg(j)} ~|~ j \in \mathbb{N}^+ \} ~\cup~ \{\text{Gap(j)} ~|~ j \in \mathbb{N}^+ \}$.
The values $j$ are used as indicators of the original DNS message sizes and inter-arrival times; higher $j$'s correlate with larger sizes and longer gaps.

\subsection{DNS Sequence Extraction}
\label{ssec:dns-sequence-extraction}
As DNS sequences only rely on coarse-grained statistical features, we can derive them even from encrypted and padded traffic.
Starting from a network traffic recording, we can first filter out non-DNS traffic.
Several identifiers show if a connection carries DNS data, such as ports (\num{853} for \acs{dot}), IP addresses (e.g., \ip{9.9.9.9}), the TLS handshake (extensions and cipher suites or \ac{sni})~\cite{Husak2015,Husak2016}, or DNS-like traffic characteristics (packets sizes of \SI{468}{\byte}).
In our implementation, we assume that we can apply an IP address and port based filter, which has proven to work perfectly for all major resolvers in practice.

Then, we reassemble the TCP streams of the identified connections and keep TLS records of type \enquote{application data}.
Processing the TLS records requires a bit of data cleanup.
Some TLS implementations use a single TLS record to fit a whole DNS packet, which makes size extraction easy.
For others, the TLS record is limited to the underlying segment size.
In this case, we can combine the sizes of consecutive TLS records if we detect that they are split at a segment boundary.
The segment boundary split can be detected by comparing the size of the TCP segment and checking if it is just below the maximum transfer unit (MTU) of the underlying IP packet.
Since TLS~1.3 the certificate is encrypted before transmission.
We need to avoid interpreting it as TLS payload, but it can easily be identified and ignored, as it is the first record after the \enquote{change cipher spec} message.
Lastly, the overhead of the TLS and potentially HTTP communication must be known, such that DNS message size extraction can correct for it.
The TLS cipher and authentication scheme are transmitted in plain text, while the HTTP overhead can be learned by sending queries to the \ac{doh} server.
At the end, we keep the message sizes and inter-arrival times of our DNS messages.

\subsection{DNS Sequence Classifier}
\label{ssec:dns-sequence-classifier}
We now use DNS sequences to build a classifier that determines if an unknown DNS sequence \enquote{equals} a previously recorded (and labeled) DNS sequence.
We leverage \acf{knn}, a simple machine learning algorithm that allows us to easily interpret and inspect the classification decisions.
To determine if two DNS sequences are the same, our classifier requires a distance function that measures how similar two objects are.
In the following, we will describe this distance function, formalize our \ac{knn} classifier and explain how it behaves for ambiguous classifications.

\subsubsection{Comparing DNS Sequences}
\label{sssec:comparing-dns-sequences}

It is important for our classifier to have a distance function which allows operating on DNS sequences with differing lengths.
One such established function is the Damerau-Levenshtein distance~\cite{Damerau1964,Levenshtein1966} or edit distance.
The Damerau-Levenshtein distance counts the number of edit operations to turn one sequence into another based on the intuition of similar sequences requiring fewer edits.
It uses four basic operations: insertion, deletion, substitution, and transposition of neighboring places.
We instantiate these four operations for both sequence element types, i.e., gaps and messages.
This way, we can transform any DNS sequence into another DNS sequence, and the distance reflects the costs of a transformation.

In its basic setting, the edit distance uses a fixed cost for each edit operation.
Unfortunately, this does not necessarily reflect the importance of multi-typed elements in our DNS sequence.
Timing information is volatile and depends on many factors outside of our control, such as network performance.
We argue that a slight difference in timing (e.g., a change $\text{Gap(2)\,}\Rightarrow\text{\,Gap(3)}$) should affect the similarity less than a full additional DNS response.
In contrast, the number of DNS queries send by the client and therefore also the number of responses should be rather constant, given the same initial configuration of the client and website.
To reflect these specifics of DNS sequences, we will assign weighted costs that depend on the elements being edited.
\Cref{tbl:distance-costs} summarizes the edit operation of our DNS sequence distance measure.
We will determine the concrete values for all cost parameters in \cref{ssec:optimizing-knn}.

\begin{table}
    \centering
    \caption{
        Summary of edit operations applicable to our DNS sequence during distance calculation.
        The \textbf{Value} column contains the concrete value, as determined through the hyper-parameter grid search in \cref{ssec:optimizing-knn}.
    }
    \label{tbl:distance-costs}
    \begin{tabular}{|l|l|c|}
        \hline
        \textbf{Operation}                    & \textbf{Cost}               & \textbf{Value}                                    \\ \hline
        Insert Msg($i$)                       & $c_{insMsg}(i)$             & $12$                                              \\ \hline
        Insert Gap($i$)                       & $c_{insGap}(i)$             & $i * 1$                                           \\ \hline
        Delete                                & Same as Insert              & Same as Insert                                    \\ \hline
        \begin{tabular}{@{}l@{}}
            Substitute \\
            Msg($i$)$\rightarrow$Msg($j$)
        \end{tabular}                         & $c_{substMsg}(i, j)$        & {\large$\frac{c_{insMsg}(i) + c_{insMsg}(j)}{4}$} \\ \hline
        \begin{tabular}{@{}l@{}}
            Substitute \\
            Gap($i$)$\rightarrow$Gap($j$)
        \end{tabular}                         & $c_{substGap}(i, j)$        & $|i - j| * 3$                                     \\ \hline
        \begin{tabular}{@{}l@{}}
            Substitute \\
            Gap($i$)$\rightarrow$Msg($j$) \\
            Msg($j$)$\rightarrow$Gap($i$)
        \end{tabular}                         & $c_{substGapMsg}(i, j)$     & $c_{insGap}(i) + c_{insMsg}(j)$                   \\ \hline
        Swap $x$ with $y$                     & $\begin{cases}
                                                     c_{swap} & x \neq y \\
                                                     0 & \text{else}
                                                 \end{cases}$               & $\begin{cases}
                                                                                   3 & x \neq y \\
                                                                                   0 & \text{else}
                                                                               \end{cases}$                                     \\ \hline
    \end{tabular}
\end{table}

\subsubsection{Formalization of \texorpdfstring{\acs{knn}}{k-NN}}
\label{sssec:formalization-of-knn}

Our \ac{knn} classifier assigns a label to unlabeled DNS sequences, based on the labels of the nearest neighbors.
This works on the assumption that sequences are similar to and have smaller distances to those traces generated from the same domain than to DNS sequences of other domains.
The parameter $k \in \mathbb{N}$ tells how many neighbors to consider.
This value is usually odd to help tie-breaking.
Our classifier is trained on a set of labeled DNS sequences $LS$.
We have a set of labels $L$, with a single label noted $l_i$, each corresponding to a website on the Internet.
A DNS sequence, named $ds_i$, is part of the set $DS$ of all DNS sequences.

We will be using two variations of \ac{knn}.
The first variant, ${knn(k, u, LS) \rightarrow (L \times DS)}$, is the traditional \ac{knn}, where given an unlabeled DNS sequence ${u \in DS}$, a set of labeled DNS sequences ${LS \subset (L \times DS): \{(l_0, ds_0), \dots, (l_n, ds_n)\}}$, and a distance function ${\mathit{dist}(ds_0, ds_1) \rightarrow \mathbb{R}_0^+}$, it returns a result set $LS_{res}$ of the closest neighbors.
The result set $LS_{res} \subseteq LS$ is a subset of $LS$ containing up to $k$ entries.
All elements in $LS_{res}$ fulfill the property, that there is no element with smaller distance, but other elements with equal distance may exist:
\vspace{-0.5em}
\begin{multline*}
\forall (l_i, ds_i) \in LS_{res}.~\\
\mathit{dist}(u, ds_i) \leq \min_{{(l_j, ds_j) \in (LS \setminus LS_{res})}} \{ \mathit{dist}(u, ds_j) \}.
\end{multline*}

The second variant, ${knn_{thres}(k, u, LS) \rightarrow (L \times DS)}$, extends the previous one, by introducing a threshold function determining if the result is sufficiently close to the search candidate.
The unconstrained (first) \ac{knn} variant may find nearest neighbors whose distance to the search candidate is significant.
This, in particular, happens if we search for DNS sequences of websites that were not part of our training phase, e.g., in an open world scenario.
In such a case, we are not interested in an arbitrary most similar result, but rather want to know whether or not the closest hits are a good classification result.
We thus define an indicator function $thres(u, ds) \rightarrow \{0, 1\}$, which returns $1$, if and only if $u$ and $ds$ are sufficiently close (otherwise, $ds \notin LS_{res}$).
The exact choice for $thres$ will be defined in \cref{sssec:scenario-2-open-world-eval}.

\subsubsection{Tie-Breaking}
\label{sssec:tie-breaking}

Regardless of the \ac{knn} variant, there is no guarantee that all labels in $LS_{res}$ are equal.
In fact, \ac{knn} can return many contradicting labels.
In this case, we perform a tie-breaking step to pick a single label out of the set of possibilities.
To this end, we first test if any label occurs in the majority and/or plurality of cases, and if so, pick this label.
Otherwise, we choose the label with the smallest distance to the search candidate under all labels which occurred most often, yet not in a plurality of cases.
We will call these \emph{pseudo-pluralities} as they are a plurality only given that their vote counts more than the votes of other equally-sized groups (due to their minimum distance).
In our evaluation, we differentiate between these tie-breaking rules.
This provides more insights into how similar the labels are, thus how well the classification works, and how much tie breaking is necessary.
Summarizing, the overall tie-breaking strategies are (from strict to flexible):
\begin{description}[noitemsep]
    \item[Unanimous:] All labels returned by \ac{knn} are identical.
    \item[Majority:] One label occurs in over half of the cases.
    \item[Plurality:] One label occurs more often than any other label but is not a majority.
    \item[Pseudo-Plurality:] Under all the labels with the highest count, choose the single label with the minimum distance to the search candidate.
    This happens if two or more labels have the same highest count, so no plurality exists.
    If multiple labels have the same smallest distance, no result can be found.
\end{description}
These tie-breaking strategies build on each other such that each \enquote{unanimous} is also a \enquote{majority} and each \enquote{majority} is also a \enquote{plurality}.
The less strict ones can assign a label, if the stricter ones cannot.

%% file: evaluation.tex
\section{Evaluation}
\label{sec:evaluation}

We will now evaluate the efficacy of our proposed methodology to classify DNS sequences that we obtain from traffic captures.
We introduce three evaluation scenarios that assess our classifier both in a closed and in an open world scenario.

\subsection{Scenarios}
\label{ssec:scenarios}
\label{ssec:datasets-generation}

In the following, we specify these scenarios, which information the attacker has, and which aspect of our implementation will be tested.
All websites described in the following scenarios are taken from the Alexa domain list~\cite{alexa} (released on 2018-11-14).

\subsubsection{Scenario~1: Closed World}
\label{sssec:scenario-1-closed-world}

The \emph{closed world} scenario is the easiest for the attacker.
Here, the attacker knows all websites that a client might visit.
Every website can be trained on and the decision only has to be which of the known websites the client visited.
This setup suites the first \ac{knn} variant (no threshold function) best, as the attacker does not run the risk of classifying non-trained samples.
Instead, the attacker can opportunistically search for the most similar DNS sequence(s).

Our closed world dataset consists of the top \num{10 000} websites.
For each website, we collected ten samples in parallel to reduce temporal effects on the samples.
We started the measurements on 2018-11-26 and they took \num{50} hours.
The remaining websites repeatedly caused errors and repeated measurements also could not succeed.
We only include traces for websites for which all ten DNS traces were retrieved successfully, which resulted in traces for 9205 websites.
Details about the measurement setup and how we handle errors are in \cref{ssec:measurement-setup}.

\subsubsection{Scenario~2: Open World}
\label{sssec:scenario-2-open-world}

The \emph{open world} scenario extends the closed world scenario in how adversaries would observe DNS traffic in practice.
Here, the attacker only knows a subset of the websites that the client visits.
This complicates the task for the attacker, as they cannot include all relevant websites in the training phase.
The attacker now not only has to find the most similar DNS sequence, but also has to decide whether this closest sequence is actually suitable.
This scenario is much more realistic, as it is impossible to determine all websites a user might want to visit.
In the open world scenario, we thus use \ac{knn} with a threshold function which can neglect results if they are \enquote{too far away} from the classification input.

To reflect an open world scenario, we split the closed world dataset in two partitions.
The top-ranked \num{1000} websites are used for training the classifier, whereas the sites with Alexa ranks \numrange{1 001}{10 000} are used for testing for false positives.
Excluding websites with erroneous traces, this results in \num{956} websites for training and \num{8249} websites for testing.

\subsubsection{Scenario~3: Cache Effects}
\label{sssec:scenario-4-cache-effects}

DNS caches drastically affect which domains need to be resolved, and thus change the DNS sequences.
We measure these \emph{caching effects} by repeatedly visiting the same website with different cache contents.
To reflect (nearly) empty caches, we only preload the TLDs and visit the websites.
To represent filled caches, we visit websites after preloading the cache with popular third-party domains.
Combining both data allows us to measure if an attacker can even fingerprint DNS contents if DNS sequences are (partially) cached.

For this scenario, we collect the top \num{10000} websites once with the normal list of only effective TLDs and once with the extended DNS cache preload list.
The extended list contains the most common third-party domains we observed in the closed world dataset.
We selected all third-party domains which were used in more than \percent{15} of domains for a total of \num{42}.
We think simulating caching through the DNS stub resolver is a fair way to simulate it, as it will remove all outgoing DNS messages.
Preloading the browser cache is harder, as frequently the same resource, like jQuery, can appear in many versions, and all of them would need to be preloaded.
We can only take into account websites that loaded successfully for both runs, leaving us with \num{8764} websites to compare.

\subsection{Measurement Setup}
\label{ssec:measurement-setup}

To bootstrap our evaluation, we created a setup to learn DNS sequences of websites.
All tests are performed on a server with a E5-2683 v4 64-core CPU with \SI{2.10}{\giga\hertz} and \SI{128}{\giga\byte} of main memory.
The server runs a Debian~9.6 with Linux~4.18.6 and is connected using a \SI{10}{\giga\byte\per\second} network interface.

There are three parties involved in DNS resolution: the browser, the local stub resolver, and the DNS resolver.
To record DNS sequences, we run 35 Docker containers in parallel, each running a Google Chrome~70 browser and an Unbound~\cite{unbound} (version~1.7.1) DNS stub resolver.
We chose Unbound, as it is already a popular choice for many Linux distributions, widely tested, and supports the dnstap~\cite{dnstap} logging functionality.
Being a stub resolver, Unbound caches DNS responses to speed up repeated queries, but will not perform the full iterative DNS lookup procedure.
Instead, it will forward all queries to an upstream recursive DNS resolver.
Here, we resort to Cloudflare's resolvers at \ip{1.0.0.1} and \ip{1.1.1.1}, given their support for \ac{dot} and documented performance~\cite{dns-perf}.

To capture a DNS sequence, we first start Unbound and Chrome.
Then, we flush and preload Unbound's DNS cache with \NS/ entries for all TLDs and effective TLDs based on the Alexa~\cite{alexa} domains we use.
Our assumption here is that a user will have these records cached due to past resolutions in the same TLD space.
Flushing the cache is important, as the Chrome start-up causes some DNS queries itself, which could alter the behavior later on.
Flushing makes the tests repeatable as we always have the same starting conditions.
We then navigate Chrome to the website we want to generate a DNS sequence for.
We let the website load for a total of 30 seconds or up to seven seconds of inactivity according to messages in Chrome's DevTool Protocol~\cite{chrome-devtools}.
Every network event (outgoing and incoming data) and script parse event (script files and \texttt{eval}) resets the inactivity timer.

To create a DNS sequence, we convert the dnstap log file into a DNS sequence by padding DNS responses to a multiple of \SI{468}{\byte}.
We label each DNS sequence with the website it was generated from.
Note that an actual on-path attacker would have to follow methodologies described in \cref{ssec:dns-sequence-extraction} to extract DNS traces from network traces.
However, we chose dnstap as the source for traces, as it greatly simplifies our experimental setups.
In fact, converting dnstap logs into DNS traces is semantically close to extracting DNS messages from the wire capture after TLS record processing.
Calculating the gap sizes and abstracted message sizes is analogous to \cref{ssec:dns-sequence-extraction}.
However, to assess potential bias, \cref{ssec:dnstap-vs-pcap} will compare the classification results from dnstap-based traces with those of traces extracted from TLS traces.

Some websites on the Alexa list only perform a redirect to other domains.
One reason are domains, which are abbreviated for quicker typing, and only redirect to the spelled out form, e.g., the American Airlines domain \texttt{\nolinkurl{aa.com}}.
Another common occurrence are websites with multiple TLDs and redirecting to the correct language version based on IP address or browser configuration.
In such cases, where an identical website is hosted under different domains, we create a single label for all of these websites.

To minimize errors in data collection, we test if a website loaded properly.
To this end, we add special marker DNS queries directly before and after loading the website, to ensure that we can reliable separate the DNS queries related to the experiment.
We then analyze the dnstap files, ensure that the marker queries are contained, and check that other DNS traffic occurred.
Using the DevTool Protocol, we filter for error messages, which indicate a DNS or connection failure.
Lastly, we check for DNS traces that are outliers when compared to other traces of the same website.
This helps to detect cases of slow network connections which led to a spurious timeout or cases where websites demanded CAPTCHA completions to continue.
In all such erroneous cases, we repeat the measurements up to two times, and discard the websites if no consistent data could be retrieved.

\subsection{Parameter Selection}
\label{ssec:optimizing-knn}

Our classifier requires a few parameter values that specify the weights for the distance function and how many neighbors should be searched for (i.e., $k$).
Choosing good constants will have a large impact on the classification results.
Consequently, we select these parameters using a hyperparameter optimization.
We choose to perform a grid search on the closed world (Scenario~1) dataset.
The dataset has ten labeled traces per websites, making 10-fold cross-validation a fitting choice.
Since the grid search is CPU heavy, and we do not want to over-fit our classifier, we select \num{5000} randomly chosen websites from the closed world dataset for cross-validation.
We choose the parameters based on the best 10-fold cross-validation results given a fixed $k=1$.
The concrete values for each edit operation are also listed in \cref{tbl:distance-costs}.
We describe the distances between traces in more detail in \cref{sec:distances-between-traces} and show how distances are distributed.

\subsection{Evaluation Results}
\label{ssec:evaluting-the-scenarios}

We will now describe the evaluation results of all scenarios that we have previously introduced.

\subsubsection{Scenario~1: Closed World}
\label{sssec:scenario-1-closed-world-eval}

The closed world scenario is the easiest of the scenarios and it will show the best case performance we can expect.
This is a fully labeled dataset which allows us to use cross-validation to measure how good our classifier is.
Concretely, we will be employing 10-fold cross-validation between all the DNS sequences.
Since we have exactly ten DNS sequences per website, we try to classify one DNS sequence based on the other nine in the training set.
This effectively limits our choice of $k$ to the range of one to nine.

\begin{figure}[t]
    \centering
    \includegraphics[width=\linewidth]{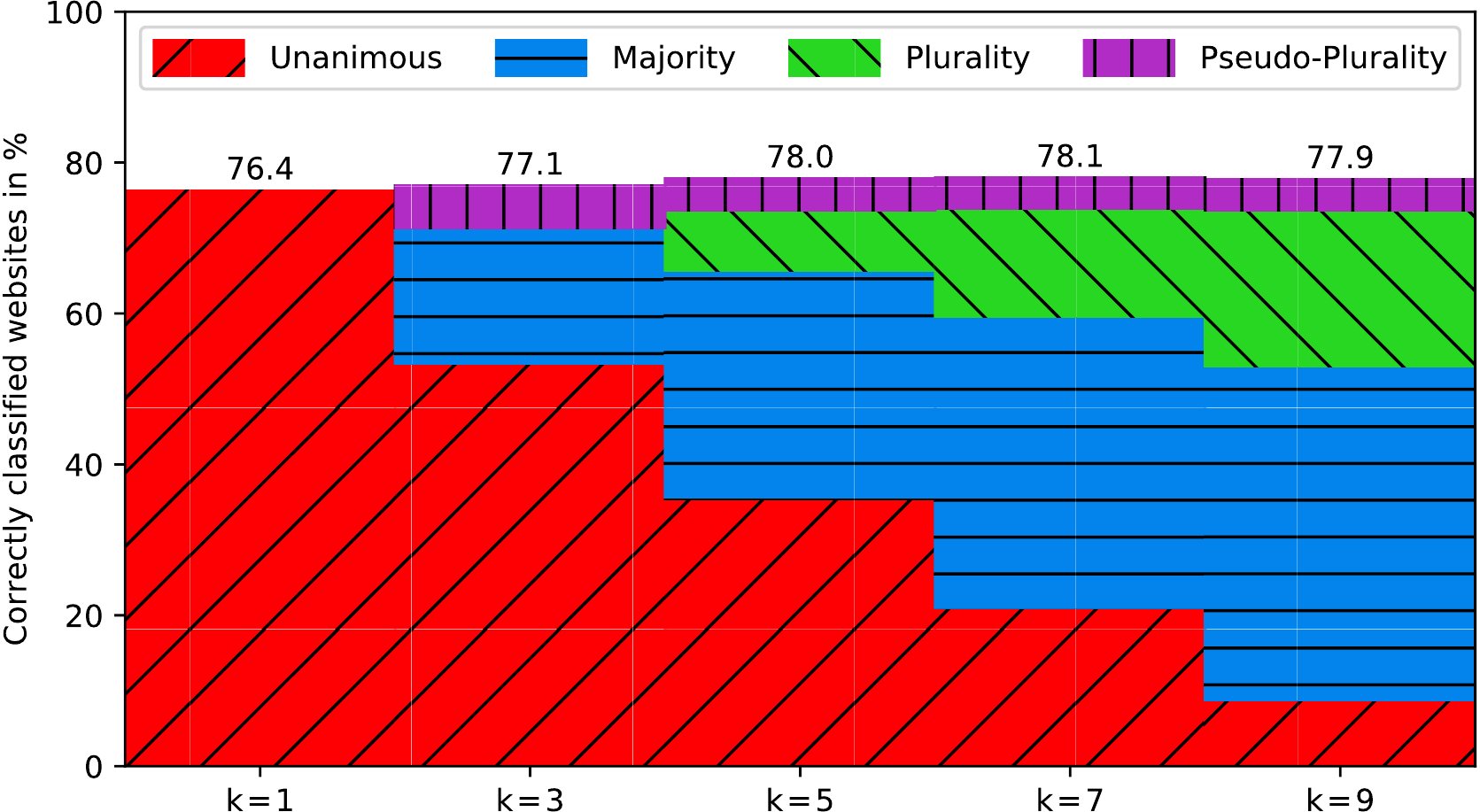}
    \caption{
        Effect of $k$ in Scenario~1 (Closed World).
        The y-axis shows the fraction of correctly-labeled DNS sequences.
        Colors denote the tie-breaking strategy.
    }
    \label{fig:s1-cw-classification-sequences}
\end{figure}

\Cref{fig:s1-cw-classification-sequences} shows the effect of $k$ on the performance of the classification.
The plot shows the percentage of correctly classified DNS sequences out of the \num{92050} sequences in total.
Irregardless of $k$, we can classify over \percent{75} of DNS sequences correctly.
Since the accuracy of even $k$'s is even worse than for odd $k$'s, such that we exclude them from further analysis.
The accuracy differs between \percent{76.4} for $k=1$ to the best $k=7$ with \percent{78.1} and is overall very stable.
We also observe that more tie-breaking is necessary for higher $k$'s.
This is intuitive, as tie-breaking only becomes important for $k > 1$.
The results also demonstrate the utility of pseudo-plurality, as majority/plurality alone does not reach the accuracy of $k=1$.

\begin{figure}[t]
    %
    \centering
    \includegraphics[width=\linewidth]{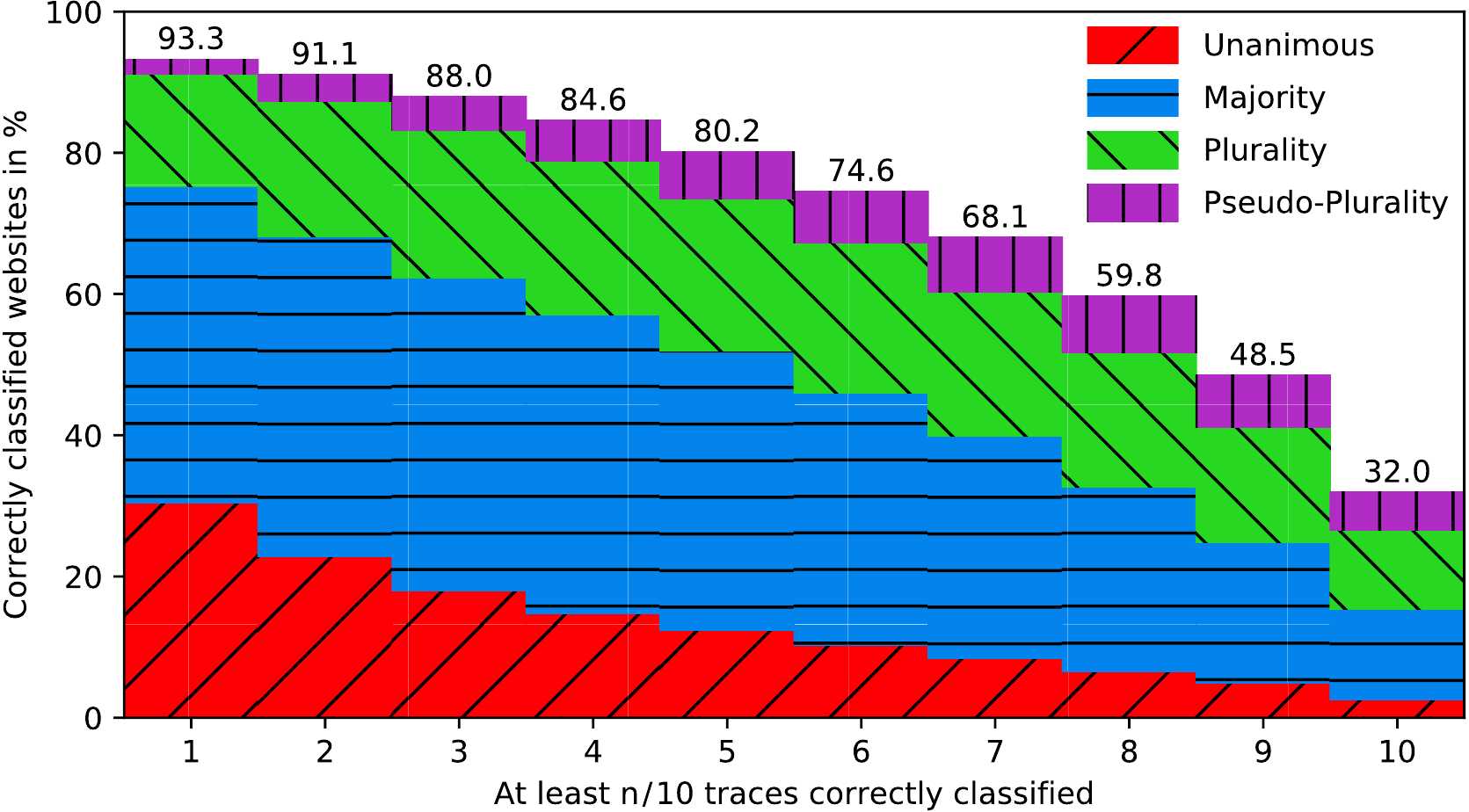}
    \caption{
        Per website results for fixed $k=7$ in Scenario~1 (Closed World).
        The y-axis shows the percentage of websites we can re-identify in how many of the 10 data points (x-axis).
    }
    \label{fig:s1-cw-classification-domains-k7}
\end{figure}

Looking from the per-website perspective, we can see that classification performs well for most of the 10 traces.
\Cref{fig:s1-cw-classification-domains-k7} shows how many of the 10 traces per website we recognize correctly for $k=7$, again broken down by tie-breaking strategy.
Our classifier provides perfect classification results (i.e., for all $10/10$ traces) for about a third of all websites.
Note that this is significantly lower than the \percent{78.1} we obtained based only on DNS sequences---a single wrong classification will decrease this number.
If we relax the ratio of correct classification per website, e.g., to \percent{80} of all DNS sequences, we can already classify about three out of five websites correctly---purely based on encrypted and padded DNS traffic.

\subsubsection{Scenario~2: Open World}
\label{sssec:scenario-2-open-world-eval}

We now turn to the open world scenario, which restricts the attacker to train only on a subset of all websites that the victim will visit.
Effectively, the open world scenario allows us to measure the \acf{fpr} of our classifier.
To obtain realistic results, we first have to take two precautions.

First, note that the training and test partitions of the dataset are disjoint.
This means that every label our classifier assigns to traces in the test dataset is a false positive.
Consequently, we turn to the second variant of our \ac{knn} classifier that foresees a threshold function $thres$ that can neglect classification candidates based on their normalized absolute distance (cf.~\cref{sssec:formalization-of-knn}).
Normalization is important, as the edit distance tends to have higher absolute values for longer sequences.
Therefore, in the threshold function, we normalize the distance by dividing through the sequence length:
$$
thres_t(u, ds) =
\begin{cases}
    1 & \frac{\text{distance}(u, ds)}{\text{max}(|u|, |ds|)} > t \\
    0 & \text{else}
\end{cases}
$$
The parameter $t$ is the threshold allowing us to change the level of our classifier's opportunism, which influences the \ac{fpr} and \ac{tpr}.
Higher thresholds favor a higher \ac{tpr}, yet will also lead to a higher \ac{fpr}.

\begin{figure}[t]
    \centering
    \includegraphics[width=\linewidth]{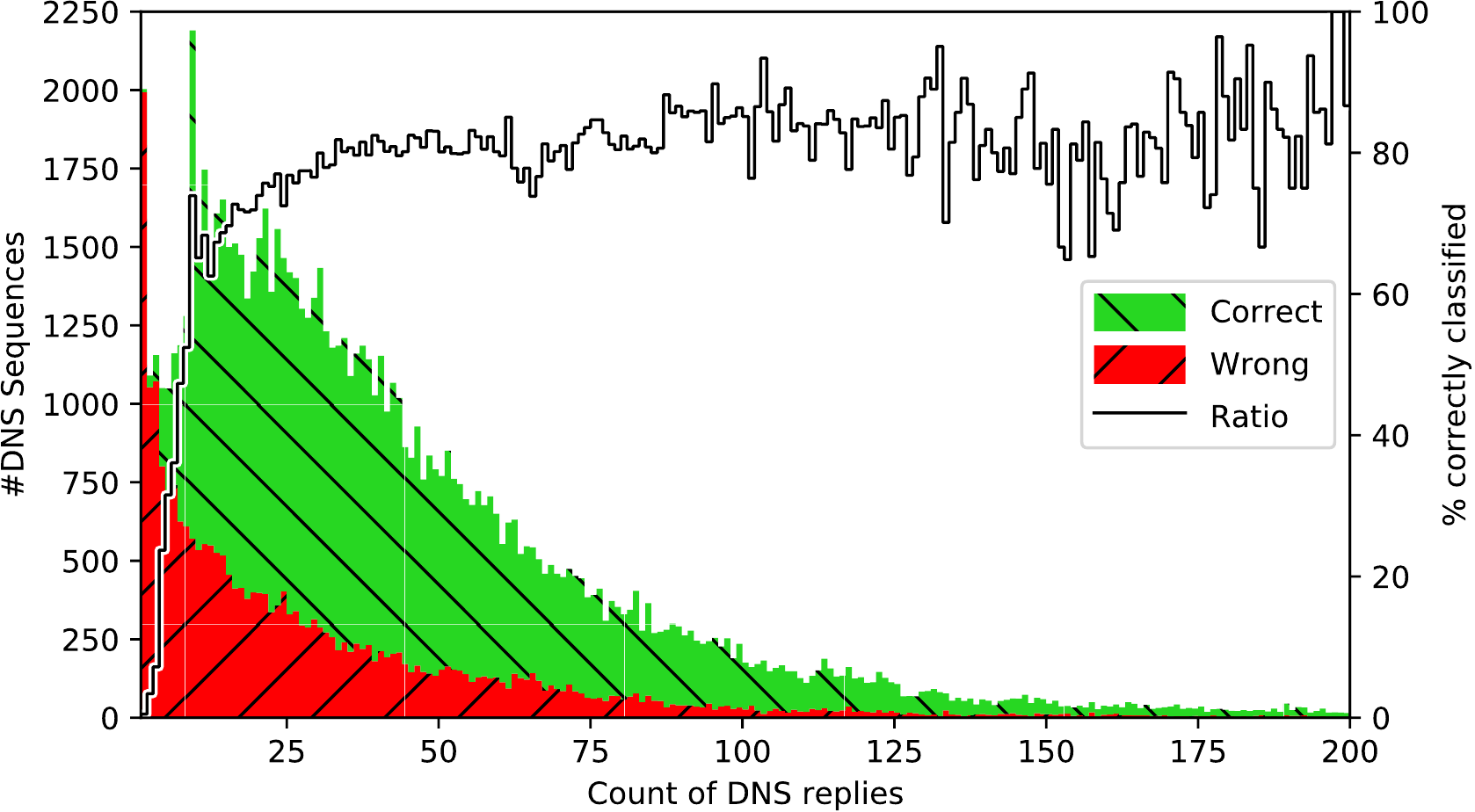}
    \caption{
        Classification results in the closed world scenario, depending on the number of Msg entries in the DNS sequence.
    }
    \label{fig:s2-sequence-entropy}
\end{figure}

Second, the open world scenario is particularly sensitive to several websites in the test dataset that provide too little entropy, as their DNS sequences are too short.
For example, a website without any third-party resources causes only one (or maximum two, in case of redirection to the \texttt{www}-subdomain) DNS request.
Given that the DNS sequences for websites that share such non-distinctive behavior look identical, it is impossible to classify these low-entropy websites correctly.
The stacked bar chart in \cref{fig:s2-sequence-entropy} illustrates this.
It reveals that the ratio of correct classifications is extremely low, and almost zero for sequences with at most three DNS messages.
The longer the DNS sequence, the higher its entropy, and the higher chances for good classification results.

An attacker can thus use the length of a DNS sequence to approximate the classification quality.
In other words, they can neglect \enquote{too short} DNS sequences, given their high risk of false classifications.
To reflect this, we regard DNS sequences up to a length of six as non-classifiable, which are all sequence lengths for which the successful classification ratio in the closed world scenario was below \percent{20} (see \cref{fig:s2-sequence-entropy}).
This affects \num{4285} (\percent{4.65507876154264}) sequences, of which \num{355} and \num{3930} sequences are in the training and test split, respectively.
We ignore them during classification and do not assign any labels, i.e., they do neither account as true positives nor as false positives.

\begin{figure}[t]
    \centering
    \includegraphics[width=\linewidth]{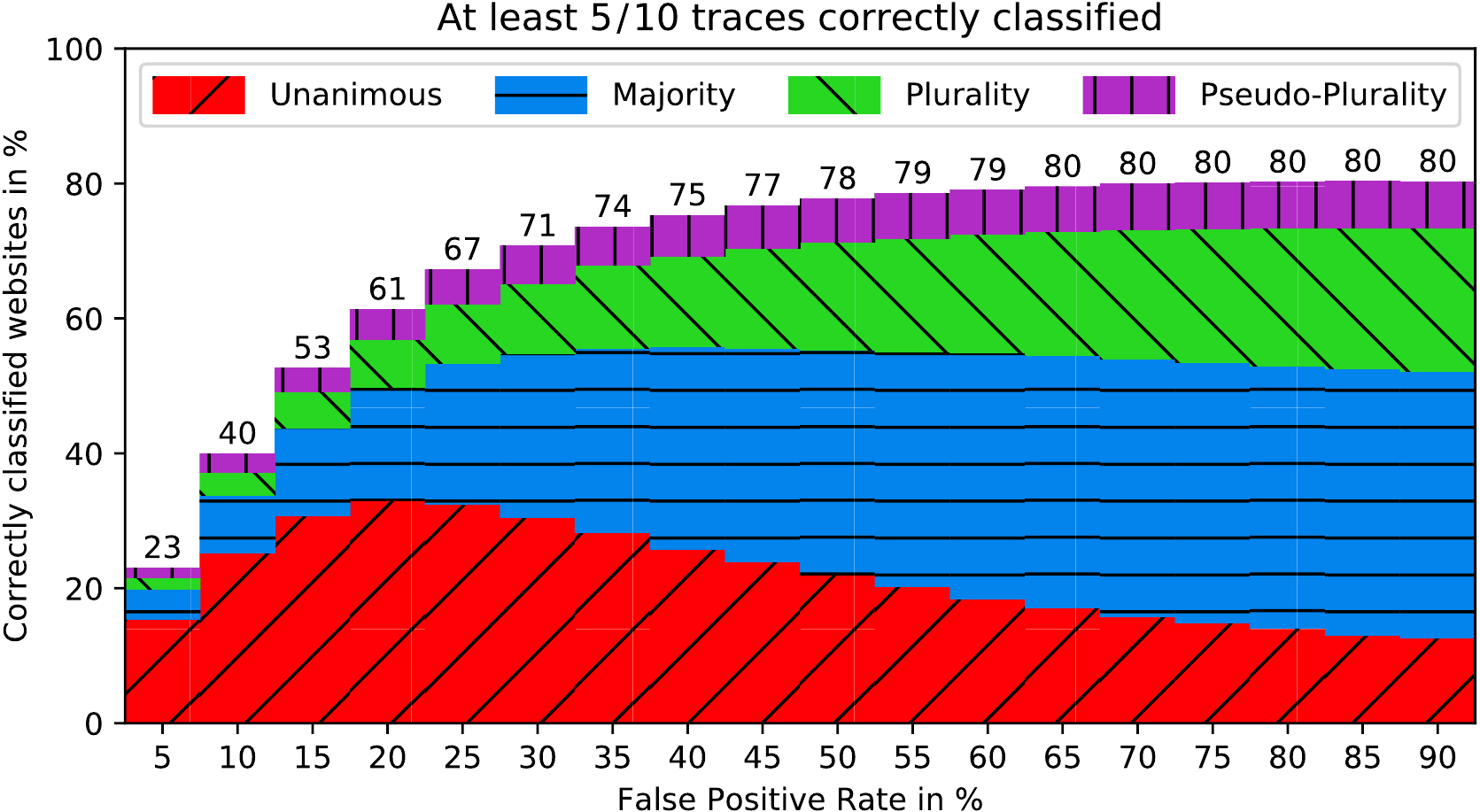}
    \caption[]{
        ROC curve showing \acs{tpr} and \acs{fpr} when varying the maximum distance threshold (Open World Scenario); $k=7$.
    }
    \label{fig:s2-effect-of-fpr-on-cw}
\end{figure}

\Cref{fig:s2-effect-of-fpr-on-cw} shows how many websites were correctly classified in the open world scenario in at least $5/10$ cases (y-axis) and relates this to the \ac{fpr} (x-axis).
We show $k=7$, as this performed best among all tested $k$ values.
The figure represents a ROC curve with a step size of \percent{5} \ac{fpr} and plots \ac{fpr} and \ac{tpr} when varying the maximum distance threshold.
If we tolerate an \ac{fpr} of \percent{10}, we can correctly classify \percent{40} of the websites in at least half of the cases.
Recall at this point that we operate on padded and encrypted data, and merely inspect the DNS traffic.
Yet, the small remaining entropy still allows us to deanonymize the browsing targets in many cases.
The \ac{tpr} even raises significantly higher for a slightly higher \ac{fpr}, e.g., \percent{61} \ac{tpr} with \percent{20} \ac{fpr}.
The lower the maximum distance threshold (i.e., the lower the \ac{fpr}), the better the classifier performs purely based on unanimous votes---spurred by the fact that the stricter threshold discards neighbors with wrong labels.
However, lower thresholds also mean that potentially correct neighbors fall above the distance threshold, which significantly lowers the \ac{tpr} compared to the closed world scenario that operated without a threshold.

\subsubsection{Scenario~3: Cache Effects}
\label{sssec:scenario-4-cache-effects-eval}

\begin{figure}[t]
\centering
\includegraphics[width=\linewidth]{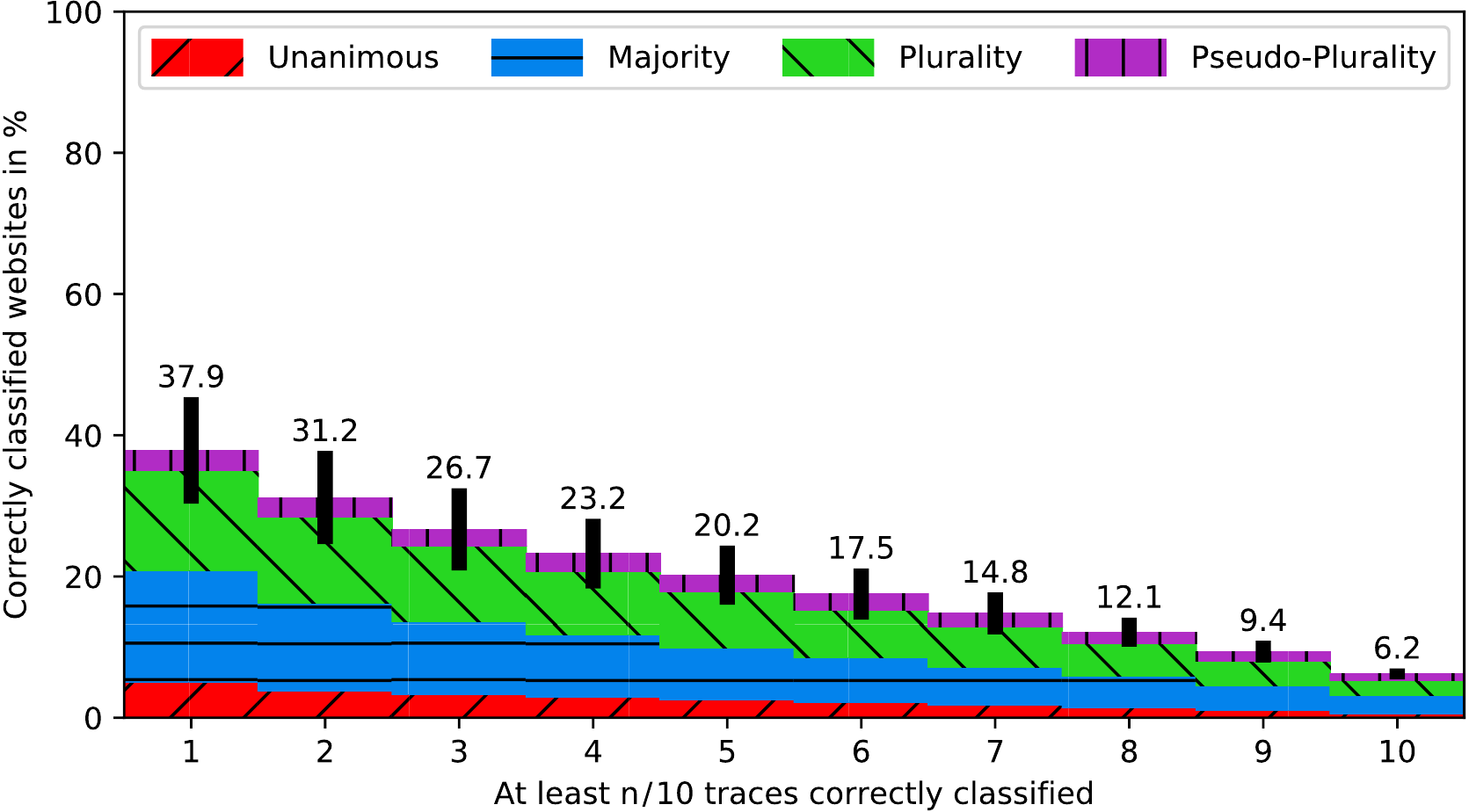}
\caption{
    Average of classifying partially cached DNS traces using a model trained on non-cached data and vice versa (Scenario~3); $k=9$.
    Error bars show the variance of results.
}
\label{fig:s4-cache-classification-domains-k9}
\end{figure}

Finally, we use the closed world scenario to evaluate if partial DNS caching will confuse a sequence-based classifier.
As described in~\cref{ssec:scenarios}, we have collected two DNS sequence datasets, one with preloaded TLDs only, and the other (\enquote{caching dataset}) with additional popular third-party domains.
In a first experiment, we will train on the non-cached dataset, and test on the caching dataset.
We repeat the experiment with training and test dataset swapped and report the average.
\Cref{fig:s4-cache-classification-domains-k9} summarizes the results for $k=9$, which fared better here.
The drop in accuracy becomes clear when comparing to the closed-world evaluation in \cref{fig:s1-cw-classification-domains-k7} that used a single cache level both for training and testing.
Only about \percent{20} of the websites were classified correctly in at least $5/10$ cases, which is far below the \percent{80} \ac{tpr} in Scenario~1.

\begin{figure}[t]
    \centering
    \includegraphics[width=\linewidth]{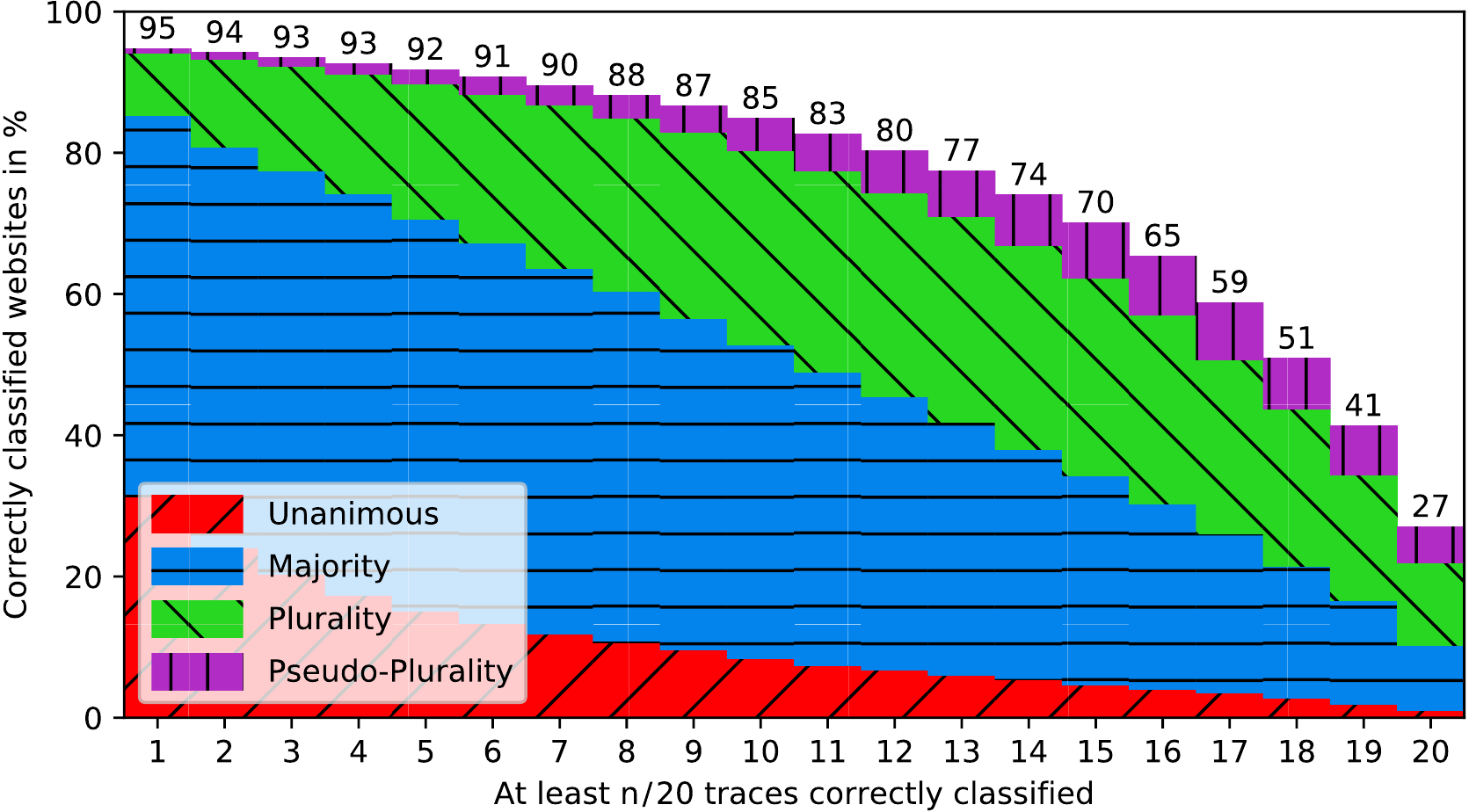}
    \caption{
        Classifying both partially cached and non-cached DNS traces using a model trained on non-cached \emph{and cached} data; $k=9$ (Scenario~3, second part).
    }
    \label{fig:s4-cache-classification-domains-k9-cross-cache}
\end{figure}

To solve this problem, an attacker can train on both a cached and a non-cached dataset.
This reflects the situation in which the attacker does not know whether a client has cached entries, and wants to be prepared for either situation.
We thus repeat the aforementioned experiment, but use a combined dataset of now 20 DNS sequences per website.
\Cref{fig:s4-cache-classification-domains-k9-cross-cache} shows the results for website classification for a fixed $k=9$ using the same 10-fold cross-validation used so far.
For the combined dataset, accuracy again increases to reasonable levels.
In fact, the results are even slightly better than in Scenario~1.
This can partially be explained with the smaller dataset size of \num{8764} compared to \num{9205} for Scenario~1, which tends to improve classification results.
Another aspect could be the fact that having more DNS sequences in the training dataset simplifies finding neighbors using \ac{knn}, as there are more correct neighbors to choose from.
Overall, the results demonstrate that caching inaccuracies can be compensated for if the training dataset combines traces obtained from both cached and non-cached website visits.

\subsection{DNS Message Trace Extraction: Dnstap vs. PCAP}
\label{ssec:dnstap-vs-pcap}

To ease our internal processing for the various experiments, we have so far extracted DNS message traces using the dnstap interface of Unbound.
However, in practice, an on-path attacker would not have this capability.
Instead, they would need to extract DNS message traces from network traffic, as described in \cref{ssec:dns-sequence-extraction}.
As a final experiment, we want to assess if message trace extraction from network traffic (in particular, PCAP traces) weakens the attacker capabilities.
We thus repeat the closed-world scenario and simultaneously record dnstap and PCAP data.
We then use each dataset to train a classifier, and finally, compare the classification results between them.

The experimental setup already foreshadows the setting of the upcoming section.
Instead of having Unbound communicate directly with Cloudflare, as described in \cref{ssec:measurement-setup}, we add two \ac{dot} proxies.
First, we add stubby~\cite{stubby} to the container, which ensures that the DNS queries follow good privacy practices.
Second, we add a tiny self-written proxy, which---for this experiment---directly forwards all messages without delaying or altering them.
In general, we will later use this second proxy to evaluate different mitigations (cf.~\cref{ssec:testing-mitigations}).
Both run inside each Docker container and listen on the loopback interface.
The PCAP data is captured before it leaves the Docker network.

\begin{table}
    \centering
    \caption{
        Classification results ($k=7$) between traces extracted from dnstap and from PCAP files.
    }
    \label{tbl:dnstap-vs-pcap}
    \begin{tabular}{ |c|*{2}{S[table-format=4.0] S[table-auto-round,table-format=2.1]|} }
    	\hline
    	\textbf{$x/10$ Traces per Domain} & \textbf{dnstap} & \textbf{in \%}    & \textbf{pcap} & \textbf{in \%}     \\ \hline
    	$1/10$                            & 8338            & 90.19904803115534 & 8229          & 89.01990480311554  \\
    	$2/10$                            & 8257            & 89.32280398096061 & 8044          & 87.01860666378192  \\
    	$3/10$                            & 8180            & 88.48983124188663 & 7828          & 84.68195586326266  \\
    	$4/10$                            & 8087            & 87.48377325832972 & 7535          & 81.51233232366941  \\
    	$5/10$                            & 7974            & 86.2613587191692  & 7188          & 77.75854608394634  \\
    	$6/10$                            & 7810            & 84.48723496321938 & 6742          & 72.9337948939853   \\
    	$7/10$                            & 7578            & 81.97749891821722 & 6112          & 66.1185633924708   \\
    	$8/10$                            & 7221            & 78.11553440069234 & 5265          & 56.955863262656855 \\
    	$9/10$                            & 6639            & 71.81955863262657 & 4158          & 44.98052790999567  \\
    	$10/10$                           & 5491            & 59.40069234097793 & 2520          & 27.260926006057986 \\ \hline
    \end{tabular}
\end{table}

\Cref{tbl:dnstap-vs-pcap} shows the comparison between the PCAP and dnstap-based classification results.
During this run, we collected traces for \num{9244} domains and again ten traces per domain.
The table compares the classification results for a fixed $k=7$ and shows how many traces per domain we could correctly classify.
The PCAP classification performs slightly worse than the dnstap results, which was expected, and can be partially attributed to our proxy setup.
Arguably, having four different programs involved in the DNS resolution process will have an impact on the stability of the timing data we can gather.
However, in general, the experiment shows that also on-path attackers that have to extract DNS traces from network traces remain powerful attackers.
In fact, the difference for domains for which at least $5/10$ traces were correctly classified is only \percent{8.5}.
While this is indeed a significant difference, in this example, on-path attackers can still classify \percent{77.8} correctly.
We thus argue that using dnstap data does not invalidate the general message of our experiments.

%% file: countermeasures.tex
\section{Countermeasures}
\label{sec:countermeasures}

In the last section we showed that attacks based solely on encrypted DNS traffic are feasible and can be successful.
We first analyze in which directions countermeasures need to be developed by understanding the impact of the two major feature types (size and timing) of our classifier.
Based on these insights, we then describe potential mitigations.

\subsection{Evaluating Perfect Mitigations}
\label{ssec:testing-perfect-mitigations}

Our classifier uses two major feature types: packet sizes and timing information.
We thus first test which feature type provides the most entropy, such that we can develop appropriate countermeasures.
To this end, we perform a thought experiment in which we assume a perfect defense for either feature type, i.e., we assume we had a perfect padding scheme and a perfect timing defense, respectively.
We simulate perfect padding by removing any Msg elements from the DNS sequences, thus removing any information about the packet sizes.
We keep Gap elements of size 0, to keep the overall message count as a feature.
Simulation of a perfect timing defense works analogously.
Here, we remove all Gap elements, which is the classifier's only source for timing information.

\begin{figure}[t]
    \centering
    \includegraphics[width=\linewidth]{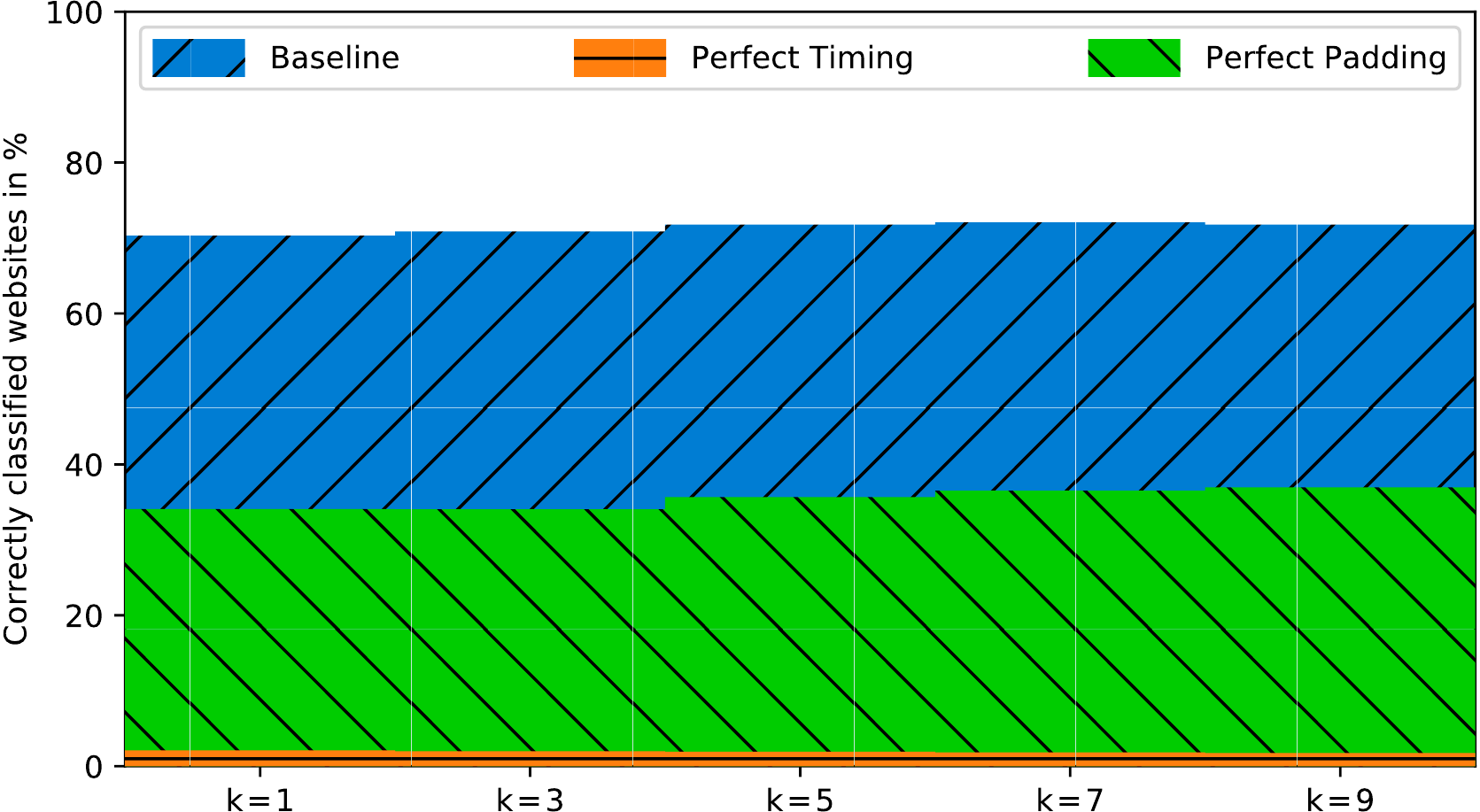}
    \caption{
        Comparison between our classifier (baseline; blue), a simulated perfect timing defense (orange), and a simulated perfect padding scheme (green).
        Numbers are not stacked.
    }
    \label{fig:countermeasures-evaluation}
\end{figure}

We re-run the closed world classification (Scenario~1) with these modifications and compare the results in \cref{fig:countermeasures-evaluation}.
The figure shows the pseudo-plurality results for all three variants, which we identified as the most promising tie-breaking strategy.
We see a large drop in performance for both simulations, which is expected, as the entropy drastically decreases.
However, when comparing both perfect mitigations, we observe a stark difference.
Most importantly, we observe that even ignoring all packet size information (i.e., perfect padding) does not entirely destroy the classifier's performance.
As expected, the \ac{tpr} significantly drops from its original level of \percent{78}, yet only to slightly below \percent{40}.
This shows that the inter-arrival timings of the DNS responses in the sequences carry significant entropy that can be leveraged to classify encrypted and perfectly-padded DNS traffic.

In contrast, a perfect timing-based defense destroys the classification results.
This illustrates that the DNS traces' entropy plummets if we remove timing information.
In fact, it also shows how little information the sizes of padded DNS responses carry.
This can be nicely illustrated by the distribution of message sizes in the closed world dataset.
An overwhelming \SI[round-precision=2,round-mode=places]{99.805}{\percent} of messages are \SI{468}{\byte} large.
Additionally, all queries were padded to the smallest size of \SI{128}{\byte}, which is why we excluded them from our feature set right away.
This situation is only slightly different if the stub resolver enables DNSSEC.
DNSSEC will trigger signature and \DNSKEY/ records that increase the size of responses such that they carry at least some entropy, i.e., in the \SIrange{936}{1404}{\byte} range.
At this point, however, we are not aware of any popular stub resolver that enables DNSSEC validation by default.
We thus chose not to report on DNSSEC-enabled numbers, although their increase in size entropy would favor our classifier.

From these observations with the proposed two optimal countermeasures we can derive important novel insights.
First, the currently proposed padding strategy~\cite{ndss-padding} is indeed a good compromise between overhead and the maximum privacy guarantees that an optimal padding could guarantee.
Yet, second, even an optimal padding strategy does not decrease the trace's entropy to a satisfactory level that preserves user privacy.
Third, countermeasures should also take into account timing information, as timing has proven to contribute significant entropy in DNS sequences.

\subsection{Evaluating Practical Mitigations}
\label{ssec:testing-mitigations}

Based on these observations, we now implement two practical mitigations and measure their efficacy and efficiency.
To this end, we select two popular defenses that hide timing information, namely a constant-rate scheme ~\cite{Dyer2012} and Adaptive Padding~\cite{Shmatikov2006}, and apply them to encrypted DNS.

\textbf{Constant-Rate Sending:} Constant-rate schemes (CR) send a packet on a fixed schedule every \SI[number-math-rm=\mathnormal, parse-numbers=false]{x}{\ms}.
The packet is filled with payload, if some is waiting, otherwise it contains padding data.
CR entirely removes the entropy in inter-arrival times, as they become constant.
However, the proposal creates significant bandwidth overhead, as every \SI[number-math-rm=\mathnormal, parse-numbers=false]{x}{\ms} a packet \emph{must} be sent, even if no payload is waiting.
Choosing a larger $x$ to reduce bandwidth overhead, however, impacts latency, as payload may have to wait for a slot before it can be sent.
Finally, CR requires a termination condition to avoid infinite transmissions of padding packets if no payload has to be sent.
This termination condition is randomized, so as not to leak how long the transmission of real payload took.
In our implementation, we define a probability $p$, which specifies how likely it is, that a dummy packet is added after the end of the stream.
This probability is sampled independently every time a dummy packet is sent.
In other words, with probability $1-p$ at each interval, we terminate the CR stream and stop sending dummy packets.

\textbf{Adaptive Padding:} Adaptive Padding (AP)~\cite{Shmatikov2006} and its improvement WTF-Pad~\cite{Juarez2016} mitigate timing side-channels by ensuring that the statistical timing features remain indistinguishable.
We focus on AP instead of WTF-Pad, as AP is a client-side only solution and does not require changes to the \ac{dot}/\ac{doh} server.
To hide timings, AP models a distribution of inter-arrival times and sends payload and padding according to the distribution.
AP uses a state machine to control when packets are transmitted.
It switches between three states:
\begin{enumerate}
    \item an \emph{idle} state in which no packets are transmitted,
    \item \emph{waiting}, representing a gap between bursts, and
    \item generate a dummy \emph{burst} with padded packets.
\end{enumerate}
A burst is a quick succession of multiple packets.
The \emph{waiting} and \emph{burst} states are controlled by statistical timing information, which need to be extracted from real traffic samples, such that the timing is realistic.
We use our closed-world dataset to extract inter-burst gaps (for \emph{waiting}) and intra-burst gaps (for \emph{burst}).
When forwarding a payload packet, AP transitions to state \emph{waiting}.
Similar to our CR implementation, AP also stops transmitting after longer times of inactivity, thus reducing the bandwidth consumed by dummy packets.
That is, with probability $p$ (as above), AP creates a dummy burst with a length chosen from the statistical distribution in the modeled data set, otherwise the state falls back to \emph{idle}.

\textbf{Comparing AP with CR:}
While both aim to hide timing information, AP and CR have different effects on the classification accuracy and result in different overheads.
We will now compare the two approaches regarding these two aspects.
To this end, we create an experimental setup that allows us to measure the schemes in varying configurations.
We first implement a \ac{dot} proxy that we place between Unbound and the \ac{dot} server, which provides these mitigations on the client side.
The proxy uses a \SI{128}{\byte} sized query which generates a \SI{468}{\byte} response for padding, since we observed this combination in \SI[round-precision=2,round-mode=places]{99.805}{\percent} of the cases.
Second, we test how much the countermeasures can mitigate privacy attacks against encrypted DNS.
Technically, we simulate their effect on our DNS sequences, which allows for rapid testing of different parameter configurations without regenerating traces.

To compare the approaches, we use our normal 10-fold cross-validation setup, and for each experiment vary the packet rate \SI[number-math-rm=\mathnormal, parse-numbers=false]{x}{\ms} (for CR) and the probability $p$ (for CR and AP).
We test CR with four packet rates of one packet every \SI{12}{ms}, \SI{25}{ms}, \SI{50}{ms}, or \SI{100}{ms}.
For both CR and AP, the probability $p$ takes values from \numrange{0.5}{0.9} in \num{0.1} increments, where larger $p$ values provide better privacy, yet worse overhead.
We measure three effects: i) the bandwidth overhead in additional DNS messages/bytes, ii) how much longer the DNS resolution process takes, and iii) how well the classification works after applying the mitigations.
We can measure the bandwidth overhead either in packets or in bytes, and we chose packets.
In our case they are basically identical, since the DNS messages are padded to mostly equal sizes.
Timing overhead only applies to CR, which can delay sending from, while AP always sends payload without any delay.
The time for the DNS resolution process is measured as the difference between the time of last DNS response and of the first DNS response, while ignoring responses created due to the dummy queries.
We ignore the responses to the dummy queries, as these do not add any overhead from a user's perspective, since these are not caused by the user/browser, but solely by the \ac{dot} proxy.

\begin{figure}[t]
    \centering
    \includegraphics[width=\linewidth]{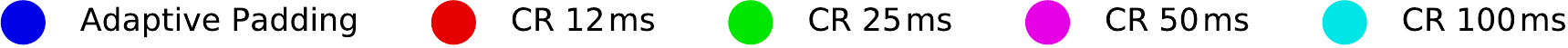}
    \includegraphics[width=\linewidth]{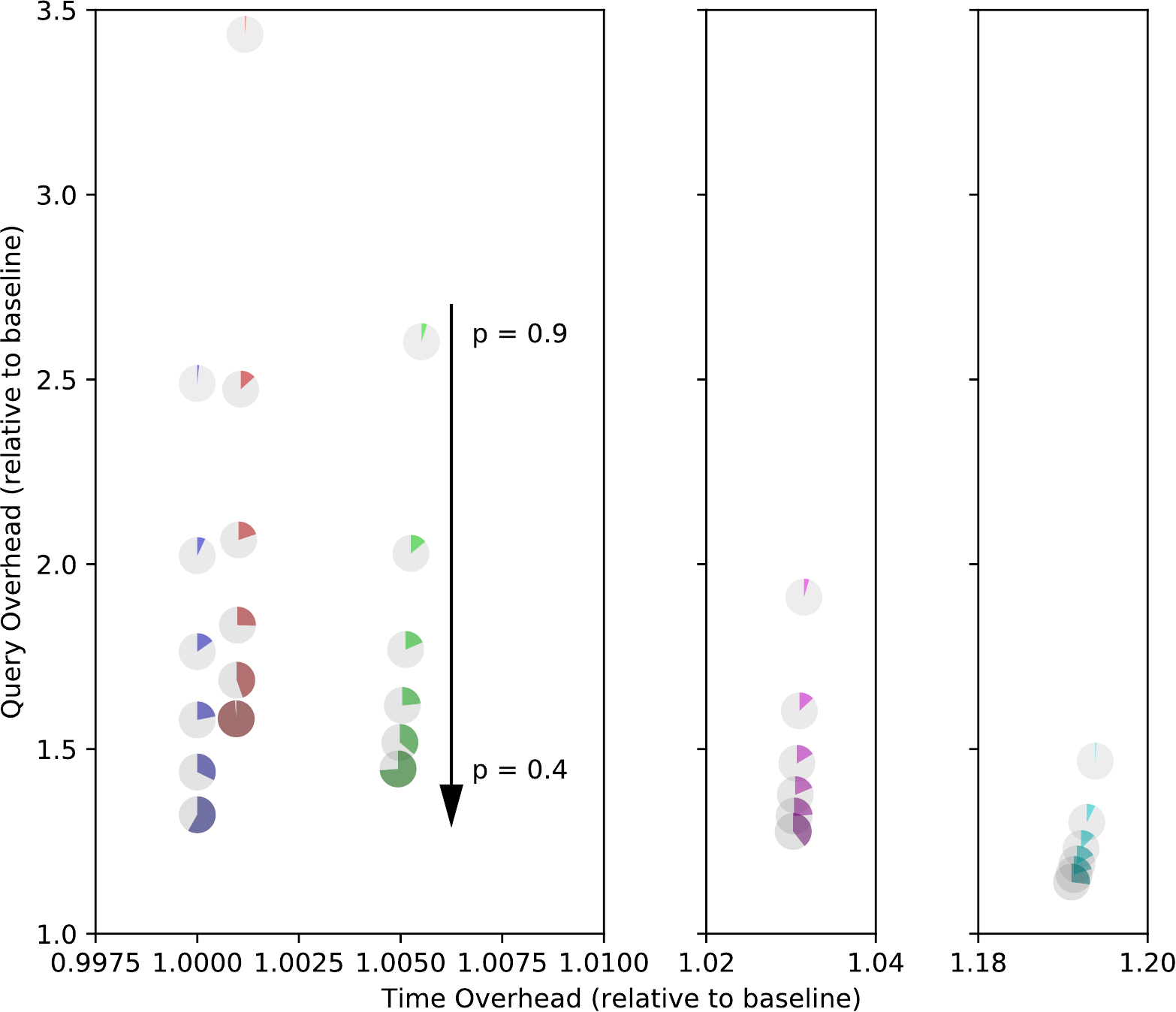}
    \caption{
        Comparison of the overhead between AP and CR.
        The x-axis shows time overhead and y-axis the packet overhead compared to the baseline at \num{1}.
        A full circle represents $10\%$ correctly labeled domains ($\geq 5/10$ traces correct).
        The \enquote{pillars} of circles correspond to the different configurations and are left-to-right in the same order as the legend.
    }
    \label{fig:mitigations-overhead-tradeoff}
\end{figure}

\Cref{fig:mitigations-overhead-tradeoff} shows the classification results.
The figure is color coded with AP being blue (left-most column of circles) and constant-rate in the remaining colors.
The x-axis shows the time overhead as a factor compared to the baseline of no modification.
A value of \num{1} indicates no overhead.
Similarly, the y-axis measures the overhead in the number of DNS messages with \num{1} being the baseline.
The size of the colored slice of each circle represents the classification results.
A full circle is equal to a \percent{10} classification success in the $5/10$ traces correctly classified setup, and smaller pie slices represent proportionally weaker classification scores.
Multiple results with varying $p$ values are directly above each other and form a \enquote{pillar} of circles.
Circles higher up in each pillar have the largest $p$ values, circles towards the bottom have the lowest $p$ values.

All variations are successful in mitigating our classifier, yet differ in the overhead they create.
In general, when comparing the two approaches at similar defense strength, AP has a higher bandwidth overhead compared to low-rate CR, while CR has a higher timing overhead.
AP can only be configured in one dimension: As expected, lowering $p$ also decreases defense strength, but improves bandwidth overhead.
By design, AP outperforms CR w.r.t.~low timing overhead, and since it gives similar privacy guarantees, can be seen as the preferred choice for devices that are not constrained in bandwidth.
For example, AP with $p=0.4$ only leaves \percent{5.819991345737776} domain classifications intact, however, incurs a query overhead of \percent{32.2001619063434}.
In contrast, the timeout-aware CR variant can be tweaked in two dimensions limit the bandwidth overhead.
Slow packet rates and low $p$ values result in relatively small bandwidth overheads, but at the trade off of larger timing overhead.
Thus, when bandwidth is of a concern, such as for mobile devices, CR with slower sending rates is probably preferred over AP.
For example, using CR with \SI{100}{\milli\second} interval and $p=0.4$ results in \percent{19.101620393816777} time overhead on average, and allows only \percent{2.7044569450454348} of the domains be classified correctly.

\subsection{Other Mitigations}
\label{ssec:realistic-mitigations}

Following our in-detail evaluation of the two selected countermeasures (AP and CR), we will now briefly describe further orthogonal countermeasures to thwart our attack.

\textbf{Timing-based Mitigations:}
A variation of the constant-rate scheme is limiting the number of unanswered queries to one and requiring that there is always a query in-flight, creating a half-duplex channel.
Walkie-Talkie~\cite{Wang2017} uses this strategy.
A half-duplex channel has advantages for HTTP communication compared to constant-rate, as the bandwidth is not limited.
This is mostly insignificant for DNS because, as we already saw, most DNS queries fit into a single IP packet and will be transferred in a single packet, thus corresponding to the constant-rate scheme with $x$ set to the round-trip time between DNS client and resolver.

\textbf{Hiding DNS traffic:}
We assumed that an attacker can identify DNS flows in traffic captures and extract DNS message sizes and inter-arrival times.
Identifying DNS flows is currently simple, as they are identifiable by their IP address, port number, or TLS handshake.
\ac{doh} has the potential to be censorship resistant, as the DNS resolver can be hosted at the same IP address as other benign content, such as a large, non-blockable \ac{cdn}.
A browser could use the same HTTPS connection to access both content and DNS.
In such a setup it is not straightforward to correctly separate the DNS from other traffic, thus preventing any DNS-based fingerprinting attacks.
Co-hosting DNS with other content is the extreme case, but teaching DNS clients to simulate browser TLS handshakes and using non-suspicious ports like \num{443} would complicate detection already.

\textbf{Morphing:}
Finally, approaches similar to traffic morphing~\cite{Juarez2016,Shmatikov2006,Wang2017,Wright2009} are conceivable also in the DNS setting.
Traffic morphing thwarts fingerprinting by transforming the statistical features of websites such that they fall into anonymity sets together with other websites, thus confusing the classifier.
In our terminology, we would aim to transform the DNS sequence of a website into the sequence of another website.
In principle, this is feasible, and an interesting direction for future work.
In fact, we could add queries to the DNS sequences (at client side), or prolong inter-arrival times (at server side)---both without causing harm to the client's functionality.
While this would add a slight communication delay, added queries could even be reasonably used to pre-fetch cached records that would expire soon.

\textbf{Private Information Retrieval (PIR)}:\acused{pir}
A stronger notion of privacy is achieved by using \ac{pir}, in which the DNS resolver is untrusted and the queries must be hidden from it.
On the one hand, our threat model assumes that the DNS resolver is trusted.
On the other hand, all schemes which achieve \acs{pir} are transferable to our setup, as the DNS resolver is always in a stronger position than a passive network attacker.
Zhao et al.~\cite{Zhao2007} try to implement a \acs{pir} scheme using \enquote{range queries}.
They analyze the privacy properties in a setup of \emph{individual queries}.
The scheme is insecure in a \emph{Web browsing} context, due to dependencies of consecutive queries, as shown by Herrmann et al.~\cite{Herrmann2014}.
Another approach is PageDNS~\cite{Asoni2017} in which DNS records are grouped into pages based on common hash prefixes or popularity of the record.
Lookups occur on a per page level and hide the exact domain.

%% file: discussion.tex
\section{Discussion}
\label{sec:discussion}

Our work is an important contribution to understand the privacy threats that DNS users face and how they can be protected.
Analyzing new side-channel attacks becomes more important, as there is a general trend for more encryption that aims to mitigate (obvious) privacy breaches.
We see this trend in the increasing adoption of HTTPS and newer encryption standards like TLS~1.3.
TLS~1.3 is the first version which encrypts the server's certificate instead of sending it in plain text.
This provides the basis to fix another domain leak in the TLS handshake due to the \ac{sni} extension, which tells the server which domain the client wants to visit.
There is ongoing work by \ac{cdn} operators and browser vendors to encrypt the \ac{sni}~\cite{ietf-tls-esni-02} and prototypes are deployed~\cite{esni-cf-firefox}.

This leaves DNS as an important target for snooping on the privacy of users.
This is because DNS is used as a first step to connect to Internet services and can thus leak the target a user visits.
Given that we constrained ourselves to encrypted \emph{and} padded DNS traffic, we find the provided classification results quite alarming.
We can partially deanonymize \percent{93.3} of websites and correctly classify \percent{78.1} of DNS sequences.
The foremost goal of our study, assessing the privacy guarantees of encrypted DNS, was thus successful, as we have shown drastic privacy problems.
The classification accuracy can be further boosted by combining our approach with existing \ac{wf} methodologies (possibly on non-DNS traffic), some of which we discuss in \cref{sec:related-work}.
Having said this, there are some limitations to our proposed methodology, which we will describe next.

\textbf{Choice of Classifier/Features:}
We used \ac{knn}, a conservative choice of a classifier, which however is easy to understand and reason about.
\ac{knn} can be slow on large training datasets, which might be prohibitive in practice.
Popular alternative approaches, such as support vector machines and neural networks, require less human feature engineering to work and might be faster during testing.

\textbf{Datasets:}
We use a rather extensive dataset with \num{9205} websites in the closed world dataset.
Related \ac{wf} attack papers regularly use a much smaller number of websites in their datasets, in the range of tens to hundreds~\cite{Hayes2016,Panchenko2016,Sirinam2018}.
A larger dataset causes precision and recall to decline~\cite[Fig.~10]{Panchenko2016}, which means our results would shine better on datasets of identical size to these papers.
In fact, we noticed this positive effect already in the evaluation of \cref{sssec:scenario-4-cache-effects-eval}.

\textbf{DNS Traffic Identification and Extraction:}
The approach to extract a DNS sequence from network captures assumes DNS servers are identifiable as such.
While \ac{dot} helps here, by specifying a dedicated TLS port, \ac{doh} uses the default HTTPS port.
So far, most \ac{doh} servers have their own dedicated IP addresses, e.g., \texttt{\nolinkurl{1.1.1.1}}.
This does not necessarily have to be the case as we already have addressed in \cref{ssec:realistic-mitigations}.

Similarly, we assumed that we can perfectly extract message sizes from the TLS/HTTPS stream.
This was made possible by the current \ac{dot}/\ac{doh} implementations, which try to use small TLS records to transmit the data, by either using a single TLS record for a DNS reply or by matching the TLS record size to the TCP maximum transmission unit.
In principle, DNS messages could be transmitted in many tiny or equally sized TLS records, hampering attempts of exact message size extraction, but increasing the cost due to more processing and network overhead.

\textbf{User Modeling:}
For the purpose of our experiments, we modeled a certain user behavior, which however might deviate in practice.
First, we assumed the client is waiting until the website has fully loaded without any background DNS traffic.
When leaving the website earlier, DNS sequences might be shorter than measured by us.
Second, most evaluations were performed with an empty browser cache and a DNS cache, which only included the effective TLDs, expect for the experiments where we determine the impact of caching.
In practice, users may have various levels of cached DNS entries, which would then result in slightly varying DNS sequences.
In addition, web resources that are shared between websites might be cached by the browser, such as jQuery or analytics libraries, which would suppress DNS queries.
While such partial caching may decrease the classifier's accuracy, one could argue that a stateful adversary can use our attack to model the DNS cache state of a user.
Knowing the (approximate) cache contents of each user, adversaries can adapt the training datasets by retraining on the traces that are expected with a certain cache state.
We plan to perform such analyses and an according adaptation to our classifier in the future.

\textbf{Lack of Entropy:}
We noticed that DNS-based classification of structurally less complex websites fails due to a lack of entropy.
This affects websites without third-party resources, but we also noticed similar problems for some \acp{cdn}.
While this is an inherent problem of DNS-based classification indeed, we argue that the continuing rise of complexity of websites will mitigate this limitation over time.

This restriction also limits the ability to transfer this attack to other domains.
Web browsing is quite unique in how it triggers diverse DNS request chains.
Other uses, such as the ones discussed in the background section (\cref{sec:background}) often cause fewer and more independent requests, which makes it hard to build DNS sequences with enough entropy.

%% file: relatedwork.tex
\section{Related Work}
\label{sec:related-work}

We now survey research that relates to the privacy of \acs{dot} and \acs{doh}, website fingerprinting, and DNS-based tracking.

\subsection{DNS Privacy}

The DPRIVE working group~\cite{ietf-dprive} aims at \enquote{providing confidentiality to DNS transactions} and thus proposed both \ac{dot} (\rfc{7858}) and \ac{doh} (\rfc{8484}).
These RFCs mention traffic analysis as a potential attack and suggest padding as a countermeasure.
Daniel Gillmor then analyzed the trade-off between padding overhead and privacy gain in the context of individual query/response pairs~\cite{ndss-padding}, which led to the padding policies (\rfc{8467}) that are the subject of our work.
We assess this general threat model to encrypted DNS in a Web setting.
We show that attackers can leverage dependencies between DNS requests to perform side-channel analysis that go beyond the transaction-wise classification by Gillmor.
In fact, we show that padding alone is insufficient to preserve privacy in encrypted DNS, which contradicts the findings and suggestions of Gillmor.
This argument is agnostic to the specific proposal.
Our work thus also affects DNS encryption schemes other than \ac{dot}/\ac{doh}.

Siby et al.~\cite{Siby2018} analyzed the privacy of \ac{doh}.
They report that their initial results indicate that domain classification in encrypted DNS is possible.
Their short paper does not provide sufficient details to allow for a full comparison with our approach.
The features are similar: size, timing, and ordering, where ordering is implicit in our (ordered) DNS sequence.
One important and crucial difference is that they do not work on padded data.
As such, their approach does not necessarily withstand existing (and deployed) countermeasures.
They report that size alone can be a distinguishing feature, whereas we show that timing has higher entropy and is vital for classification of padded encrypted traffic.

Shulman studied the \enquote{pitfalls of DNS encryption}~\cite{Shulman2014}.
She analyzed DNSCurve~\cite{dnscurve} and DNSCrypt~\cite{dnscrypt}, two DNS over TLS proposals (both different from \ac{dot}) and opportunistic IPsec.
In a strict threat model, Shulman shows that an indistinguishability privacy definition is unachievable.
If the attacker can select two domains, one of which the client requests, and can monitor the authoritative servers, it is easy to detect which domain the client picked.
Any side channel, be it IP address, size, or response time, can be used to differentiate.
Since the attacker controls the domains, they can always choose a pair of domains which are distinguishable.
In contrast to our threat model, Shulman follows an overly strict privacy definition.
Allowing the attacker to choose the domains is a very strong assumption and, in a fully encrypted Web, unrealistic.
Furthermore, the ability to differentiate between two attacker-controlled domains is not particularly useful for an attacker.
In contrast, our goal, i.e., identifying which websites users visit, is the core of user profiling.

\subsection{\acl{wf} Attacks}
\label{ssec:website-fingerprinting-attacks}
In contrast to DNS traffic analysis, \ac{wf} is a well-established research field.
A wide range of proposed classifiers successfully reveal the websites from either encrypted traffic or from traffic of anonymous communication networks (such as Tor), including Support Vector Machines (SVM)~\cite{Panchenko2016,Panchenko2011}, random forests~\cite{Hayes2016}, \ac{knn}~\cite{Wang2014}, and more recently, neural networks~\cite{Hayes2016,Schuster2017,Sirinam2018}.
The underlying features of these approaches are often similar and include, among other things, packet counts/sizes per direction, direction changes, packet orderings, bursts, and inter-arrival times.
Our \ac{knn} classifier borrows several features from the \ac{wf} domain for a new context.
Yet, in contrast to \ac{wf} approaches, we have to restrict ourselves to DNS flows, which are significantly less rich than HTTP(S) flows.
In fact, one can argue that DNS transactions typically trigger multiple HTTP transactions, and such transactions (even if padded) have significantly higher entropy than DNS messages padded to quantized size.

Greschbach et al.~\cite{Greschbach2017} use DNS to improve deanonymization attacks against Tor users.
Their proposal combines a \ac{wf} attack on the ingress node with capturing DNS messages on Tor exit nodes.
They measure how many autonomous systems a query traverses to the DNS resolver, and determine that the safest DNS option for exit nodes is their \acposs{isp} DNS resolver.

To the best of our knowledge, we are the first to apply the idea of sequence-based traffic analysis to deanonymize DNS traffic.
Indeed, \emph{plaintext} DNS sequences have been used by Wang et al.~\cite{Wang2019} as features to perform \ac{wf}.
However, they assume access to the content of DNS queries, which allows for trivial website inferences.
We follow a much stricter attacker model and infer the website from encrypted and padded DNS traces by investigating timing-based and statistical features.

\subsection{DNS-Based User Tracking}
\label{ssec:tracking-with-dns}
Non-encrypted DNS queries have also been abused to track users, e.g., by Herrmann et al.~\cite{Herrmann2013} and Kirchler et al.~\cite{Kirchler2016}.
They analyze how a passive adversary that intercepts traffic between the user and DNS resolver or the DNS resolver itself can track individual users even though their IP addresses change.
They show how behavior patterns of users can be used to link user sessions with a high accuracy even over longer periods of time.
Our work is largely orthogonal to these papers, since we assume that the DNS resolver is trusted and that the connection between client and DNS resolver is encrypted.

%% file: conclusion.tex
\section{Conclusions}
\label{sec:conclusions}

Our work underlines the importance of carefully studying the possibility of traffic analysis against encrypted protocols, even if message sizes are padded.
While there is a plethora of literature on \acl{wf} based on HTTPS and Tor traffic, we turned to encrypted DNS---an inherently more complex context, given the low entropy due to short sequences and small resources.
To the best of our knowledge, we are the first to show that passive adversaries can inspect sequences instead of just single DNS transactions to break the widely deployed best practice of DNS message padding.
We plan to share these novel findings with the respective working group at the IETF to help standardize effective countermeasures that go beyond the known transaction-based traffic analysis attacks.
We hope that our observations will foster more powerful defenses in the DNS setting that can withstand even more advanced traffic analysis attacks like ours.

%% file: appendix.tex
\begin{appendix}

\section*{Distances between Traces}
\label{sec:distances-between-traces}

\begin{figure}[h]
    \centering
    \includegraphics[width=\linewidth]{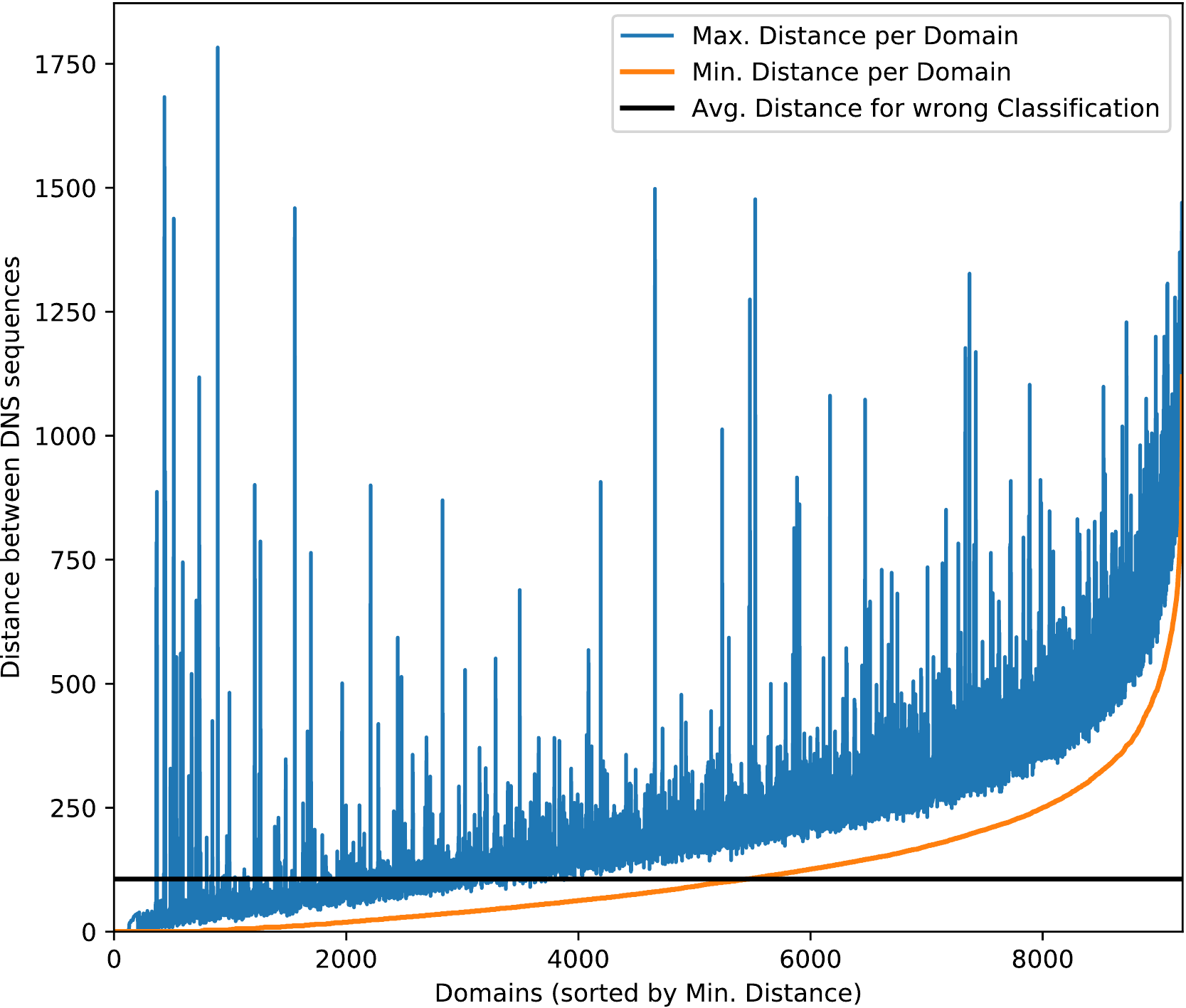}
    \caption{
        Overview over the distances between traces of the same domain.
        The traces are from the closed-world scenario.
        The orange line shows the minimal distance between all 10 traces of a domain, while the blue line shows the maximal distance.
        The black line displays the average of the minimal distances for wrong classifications.
    }
    \label{fig:distance-per-cluster}
\end{figure}

\begin{figure}[h]
    \centering
    \includegraphics[width=\linewidth]{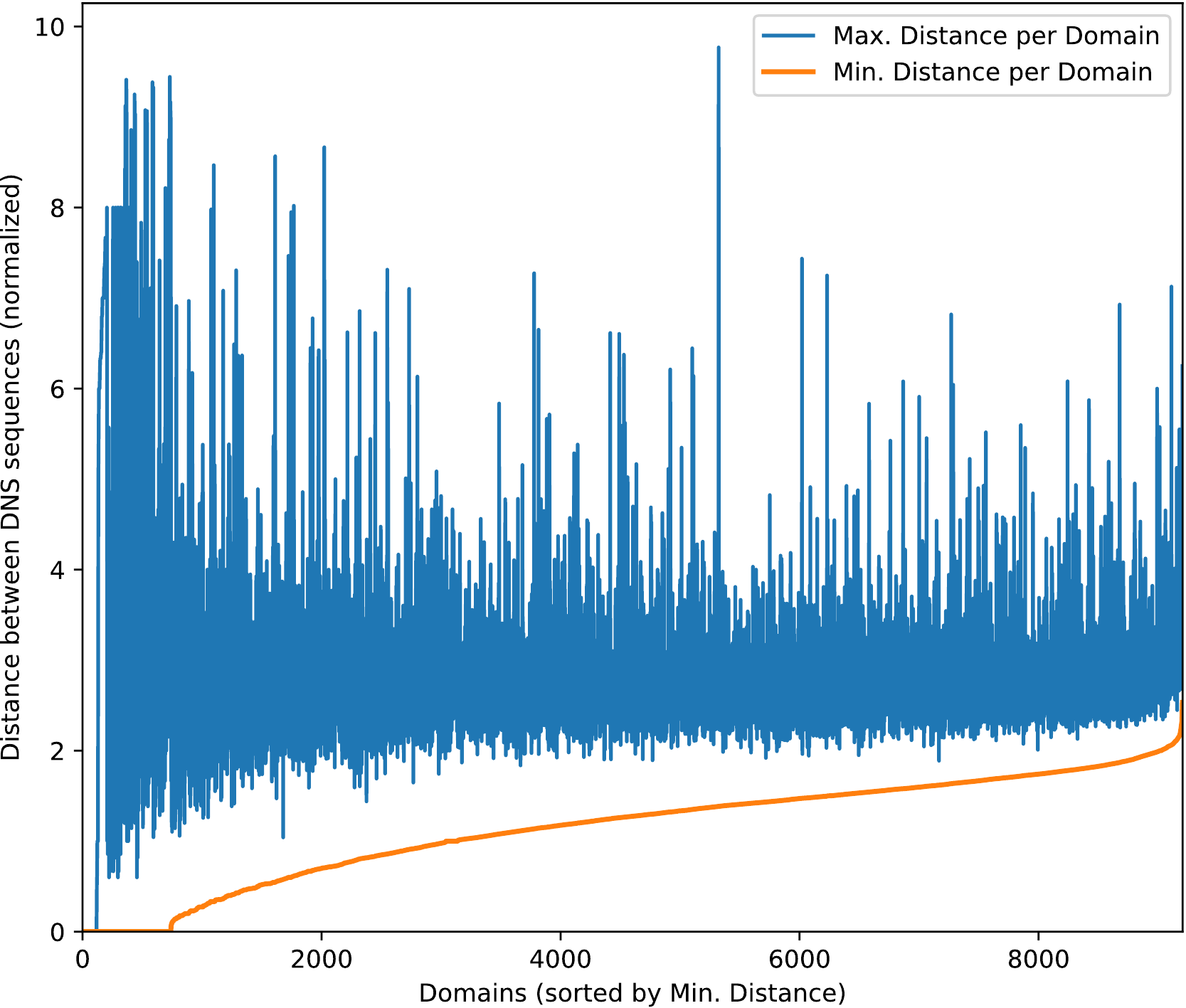}
    \caption{
        Analogue to \cref{fig:distance-per-cluster} but with normalized distances.
        Distances are normalized by dividing the distance value by the length of the longer sequence.
    }
    \label{fig:distance-per-cluster-normalized}
\end{figure}

\newpage
In \cref{ssec:optimizing-knn} we determined concrete values for our distance function.
\Cref{fig:distance-per-cluster} shows the distance function for the domains in our closed-world dataset, as specified in \cref{sssec:scenario-1-closed-world}.
It shows the minimal (orange) and the maximal (blue) distance between all ten traces of a domain.
The domains are sorted by the minimal distance.
The black line shows the average of the minimal distance for wrongly classified traces.

From the graph we can see that we have a set of domains with identical traces.
There are \num{118} domains where all traces are identical and \num{739} with at least two identical traces.
The distances largely depend on the length of the sequence, which is a consequence of our choice of using the Damerau-Levenshtein distance.

\Cref{fig:distance-per-cluster-normalized} shows the same as before, but with normalized distances.
We normalize the distance by dividing the value by the length of the longer sequence as described in \cref{sssec:scenario-2-open-world-eval}.
Here, we can see that the distances are much more uniform.
The quick increase of distance values towards the right side of \cref{fig:distance-per-cluster} is gone, which indicates that these large distances belong to very long sequences and thus complex websites.

The average distance for wrong classifications is quite low, since there are many wrong classifications with a distance of exactly zero.
This is a problem with simple websites without many third-party resources, such that our DNS sequences are small and identical across multiple domains.
We address this more in our discussion section (cf. \cref{sec:discussion}).

\end{appendix}